\documentstyle{elsart}
\input epsf

\begin{document}

\begin{frontmatter}

\hfill INP 1798/PH

\vspace{5mm}

\title{Collective modes and current-algebraic sum rules in nuclear medium%
\thanksref{grants}}
\thanks[grants]{Research supported by PRAXIS~XXI/BCC/429/94,
        the Polish State Committee for
        Scientific Research grant 2P03B-080-12,
        PCERN/FIS/1062/97 and PESO/1057/95}

\thanks[emails]{\hspace{0mm} E-mail addresses: broniows@solaris.ifj.edu.pl,
 brigitte@hydra.ci.uc.pt}

\author[INP]{Wojciech Broniowski} and
\author[Coimbra]{Brigitte Hiller}

\address[INP]{H. Niewodnicza\'nski Institute of Nuclear Physics,
         PL-31342 Krak\'ow, Poland}

\address[Coimbra]{Centro de {F\'{\i}sica T\'eorica}, 
University of Coimbra, P-3000 Coimbra, Portugal}

\begin{abstract}
In-medium sum rules following from the chiral
charge algebra of QCD are reviewed, and new sum rules are derived.
The new sum rules relate the $I^G(J^{PC})=1^-(0^{++})$ excitations
(quantum numbers of $a_0(980)$)
to the scalar and isovector densities, and are nontrivial for the
isospin-asymmetric medium. We present an extensive illustration of the sum
rules with help of quark matter in the Nambu-Jona--Lasinio model.
Collective excitations different from the usual meson branches
(spin-isospin sound modes) are shown
to contribute significantly to the sum rules and to play a crucial role
in the limit of vanishing current quark masses.
\end{abstract}


\end{frontmatter}



\section{Introduction}

\label{sec:intro} Over the past years intense efforts have been made to
better understand the properties of nuclear systems under extreme conditions
\cite{heidelberg,hadrons}. It is commonly accepted that basic properties of
hadrons undergo severe modifications in nuclear medium \cite
{Brown:mat,Weise:mat,Adami,walecka,Celenza:mat,%
brscale,Birse:rev,BrownRho:physrep}
We expect that at sufficiently large densities chiral symmetry is
restored. Moreover, we know that already at nuclear saturation density we
should find strong medium effects. For instance the quark condensate $%
\langle \overline{q}q\rangle $ is estimated to drop to about 70\% of its
vacuum value at the nuclear saturation density, as follows from the
model-independent prediction of Refs. \cite{druk,grieg}. The change in this
basic scale of strong interactions, as well as other matter-induced effects,
undoubtedly lead to severe modifications of in-medium hadron properties,
whose excitation energies, widths, coupling constants, size parameters, {\em %
etc.} undergo changes. The experimental evidence for these effects can be
found in studies of mesonic atoms, or in the measurements of dilepton
spectra in heavy--ion collisions in the Ceres \cite{ceres} and Helios \cite
{helios} experiments at CERN. Much more accurate data on hot and dense
matter will be provided by the Hades experiment, and by RHIC in the near
future. It is therefore an important task to better understand and describe
theoretically {\em mesonic excitations} in dense and hot systems.

Recent years have brought new interesting ideas and developments in this
field. The incomplete list, relevant for the subject of this paper, contains
the possibility of S-wave kaon condensation in nuclear matter \cite
{kaon,PolWise,Brown:kaon,Lutz}, and the application of chiral effective
Lagrangians and models \cite
{Lutz,Hatsuda85,BMZ:temp,ReinDang,Vogl1,Lutz92,Jaminon2,RuivoHiller,%
Klevansky}
to nuclear systems. General model-independent predictions for excitations
with quantum numbers of the pion, based on chiral charge algebra, were made
in Refs. \cite{Lutz,tdcwb:gb,tdcwb:isovector,vsum,algnjl}. Our present
work summarizes and further extends the results presented there.

The purpose of this paper is twofold: In the first part we review the
previously-derived current-algebraic sum rules for pionic excitations in
nuclear medium (the generalization of the Gell-Mann--Oakes--Renner relation
\cite{Lutz,tdcwb:gb}, the sum rule of Ref. \cite{tdcwb:isovector}), as well
as derive new sum rules concerning the excitations with quantum numbers of
the $\delta $ (or $a_0)$ meson ($I^G(J^{PC})=1^{-}(0^{++})$) (Sec. \ref
{sec:current}). We discuss formal predictions following from these sum rules
(Sec. \ref{sec:formal}). Particular attention is drawn to nuclear matter
with isospin asymmetry, since this is the case where nontrivial conclusions
can be drawn for the behavior of mesonic excitations in the limit of
vanishing current quark masses. We discuss the appearance of very soft modes
in this limit. In the pion channel there exists a positive-charge mode (for
medium of negative isospin density) whose excitation energy scales in the
chiral limit as the current quark mass itself, and the the square root of
it, as is the case of the vacuum. In the $\delta $ channel there exists a
positive-charge mode (for medium of negative isospin density) whose
excitation energy scales as the difference of the current masses of the $u$
and $d$ quarks. These modes are shown to completely saturate the sum rules
in the limit of vanishing current quark masses.

In the second part of the paper (Sec. \ref{sec:njl}-\ref{sec:asymm}) we
present an extensive illustration of the general results with help of quark
matter in the Nambu-Jona--Lasinio model \cite{njl}.
Although quark matter is not a
realistic approximation to nuclear matter (except, perhaps, at very large
densities), the model is good for the present purpose. The reason is that
the Nambu-Jona-Lasinio model is consistent with chiral symmetry and complies
to chiral charge algebra relations leading to the sum rules. We show that
the results of the model are highly nontrivial: collective states appear in
isospin-asymmetric medium (spin-isospin sound modes) and these states are
necessary to saturate the sum rules. For certain choice of model parameters,
these sound modes become the very soft modes in the limit of vanishing
current quark masses, and they completely saturate the sum rules. Finally,
we remark that the Nambu--Jona-Lasinio
model is interesting in its own, and that much of the
expectations concerning the behavior of mesons in medium have been based on
calculations carried out in this model \cite
{Lutz,Hatsuda85,BMZ:temp,ReinDang,Vogl1,Lutz92,Jaminon2,RuivoHiller,%
Klevansky,Bijnens:review}.

\section{Current-algebraic sum rules}

\label{sec:current} In this section we present a set of sum rules that are
going to be explored in this paper. The method follows Refs. \cite
{tdcwb:isovector,algnjl}. The sum rules follow from the $SU(2)\otimes SU(2)$
chiral charge algebra \cite{current:algebra:1,current:algebra:2} of QCD and
involve no extra assumptions, therefore are very general. In the context of
effective chiral models such relations were derived in Refs. \cite{Lutz}. In
this section we also derive the corresponding relations involving the vector
current, {\em i.e.} involving the excitations with quantum numbers of the $%
\delta $ (or $a_0(975)$) meson, with $I^G(J^{PC})=1^{-}(0^{++})$. For the
simplicity of notation the derivation is made for two flavors, generically
denoted by $u$ and $d$. The cases involving strangeness ($K$ and $K_0^{*}$
excitations) can be obtained from the results below by replacing $u$ or $d$
by $s$.

\subsection{Operator identities}

\label{sec:ids} Consider the charges corresponding to vector and axial
vector rotations, defined in the usual way as $Q^a=\int d^3xJ_0^a$ and $%
Q_5^a=\int d^3xJ_{5,0}^a$, with the appropriate currents defined as $J_\mu
^a=\overline{\psi }\gamma _\mu \frac{\tau ^a}2\psi $ and $J_{5,\mu }^a=%
\overline{\psi }\gamma _\mu \gamma _5\frac{\tau ^a}2\psi $. The charges
satisfy the $SU(2)\otimes SU(2)$ chiral charge algebra: 
\begin{equation}
\lbrack Q^a,Q^b]=i\epsilon ^{abc}Q^c,\;\qquad [Q_5^a,Q_5^b]=i\epsilon
^{abc}Q^c.  \label{su2xsu2}
\end{equation}
The density of the QCD Hamiltonian is denoted as ${\cal H}_{{\rm QCD}}$. We
need explicitly the mass term, ${\cal H}_{{\rm mass}}=\overline{\psi }{\cal M%
}\psi $, where the current mass matrix is ${\cal M}={\rm diag}(m_u,m_d)$.
The canonical anticommutation rules for the quark operators, $\{\psi _\alpha
(\vec x,t),\psi _\beta ^{\dagger }(\vec y,t)\}=\delta ^3(\vec x-\vec
y)\delta _{\alpha \beta }$, and the explicit form of ${\cal H}_{{\rm mass}}$
result in the following operator identities:

\begin{eqnarray}
\lbrack Q^a,[Q^b,{\cal H}_{{\rm QCD}}(0)]] &=&\overline{\psi }(0)[\tau
^a/2,[\tau ^b/2,{\cal M}]]\psi (0),  \label{commV} \\
\lbrack Q_5^a,[Q_5^b,{\cal H}_{{\rm QCD}}(0)]] &=&\overline{\psi }(0)\{\tau
^a/2,\{\tau ^b/2,{\cal M}\}\}\psi (0).  \label{commA}
\end{eqnarray}
Rewriting these relations for the neutral and charged channels (with $\tau
^{\pm }=\left( \tau ^1\pm i\tau ^2\right) /\sqrt{2}$), we obtain the
following operator identities: 
\begin{eqnarray}
\lbrack Q_5^0,[Q_5^0,{\cal H}_{{\rm QCD}}(0)]] &=&m_u\overline{u}u(0)+m_d%
\overline{d}d(0),  \label{commA0} \\
\lbrack Q_5^{-},[Q_5^{+},{\cal H}_{{\rm QCD}}(0)]] &=&\frac
12(m_u+m_d)\left( \overline{u}u(0)+\overline{d}d(0)\right) ,  \label{commA1}
\\
\lbrack Q^0,[Q^0,{\cal H}_{{\rm QCD}}(0)]] &=&0,  \label{commV0} \\
\lbrack Q^{-},[Q^{+},{\cal H}_{{\rm QCD}}(0)]] &=&\frac 12(m_u-m_d)\left( 
\overline{u}u(0)-\overline{d}d(0)\right) .  \label{commV1}
\end{eqnarray}
Relation (\ref{commV0}) is trivial, since the third component of isospin is
a good symmetry even when $m_u\neq m_d$. Relation (\ref{commV1}) is
nontrivial only if $m_u\neq m_d$.

\subsection{Gell-Mann--Oakes--Renner relations in medium}

\label{sec:GMOR} In Appendix \ref{app:sr} we present a detailed derivation
of sum rules from the above operator identities, and the reader is referred
there for the details. The sum rules are obtained by the usual technique:
Identities (\ref{commA0}--\ref{commA1}, \ref{commV1}) are sandwiched by a
state $|C\rangle $, given below. Then, a complete set of intermediate
states is inserted in the LHS of the identities.

The state $|C\rangle $ is chosen to be a uniform, translationally--invariant
state describing nuclear matter. It has fixed baryon number density, $\rho _B
$, and isospin density, $\rho _{I=1}$. We choose to work in the rest frame
of nuclear matter. Let us explain the notation used below: states $%
|j^a\rangle $, where $a$ labels isospin, denote all states that can be
reached from the state $|C\rangle $ by the action of the appropriate
current. For instance, in the
case of the $J_{5,0}^0$ operator, the states $%
|j^0\rangle $ have quantum numbers of the neutral pion, and include all
possible modes excited ``on top'' of nuclear matter: the vacuum pion branch,
collective modes, 1p-1h, 2p-2h, {\em etc.}, excitations of the Fermi sea, 
{\em etc.} As shown in Appendix \ref{app:sr}, the sum rules involve
intermediate states with momentum $0$ in the nuclear matter rest frame. The
quantity $E_{j^a}$ denotes the excitation energy of the state $|j^a\rangle $
(in the rest frame of nuclear matter). The symbol $\sum_{j^a}$ includes the
sum over discrete states, as well as the integration over continuum states.

Relations (\ref{commA0}-\ref{commA1}) result in sum rules which are
in-medium generalizations of the Gell-Mann--Oakes--Renner (GMOR) relations 
\cite{GMOR}:

\begin{eqnarray}
-m_u\langle \overline{u}u\rangle _C-m_d\langle \overline{d}d\rangle _C
&=&\sum_{j^0}\left| \langle j^0|J_{5,0}^0(0)|C\rangle \right| ^2,
\label{GMOR0} \\
-(m_u+m_d)\langle \overline{u}u+\overline{d}d\rangle _C &=&\sum_{j^{-}}{\rm %
sgn}(E_{j^{-}})\left| \langle j^{-}|J_{5,0}^{-}(0)|C\rangle \right| ^2 
\nonumber \\
&&+\sum_{j^{+}}{\rm sgn}(E_{j^{+}})\left| \langle
j^{+}|J_{5,0}^{+}(0)|C\rangle \right| ^2.  \label{GMOR1}
\end{eqnarray}
Indeed, in the case of the vacuum, $|C\rangle =|{\rm vac}\rangle $, we can
single out the one-pion contribution in the RHS of Eq. (\ref{GMOR0}-\ref
{GMOR1}). Let us denote this state (with 3-momentum 0) as $|\pi ^a\rangle $.
For example, in the case of (\ref{GMOR0}) we then find $\langle \pi
^0|J_{5,0}^0(0)|{\rm vac}\rangle =m_{\pi ^0}F_{\pi ^0}$, where $m_{\pi ^0}$
and $F_{\pi ^0}$ are the neutral pion mass and decay constant. Therefore, we
can write 
\begin{equation}
-m_u\langle \overline{u}u\rangle _{{\rm vac}}-m_d\langle \overline{d}%
d\rangle _{{\rm vac}}=m_{\pi ^0}^2F_{\pi ^0}^2+\sum_{j^0\neq \pi }\left|
\langle j^0|J_{5,0}^0(0)|C\rangle \right| ^2=m_{\pi ^0}^2F_{\pi ^0}^2+{\cal O%
}(m^2).  \label{GMORvac}
\end{equation}
The symbol $\sum_{j^0\neq \pi }$ denotes the sum over all contributions
other than the one-pion-state, {\em e.g.} three pions, $\rho +\pi $, {\em %
etc. }As is well known, such contributions are chirally suppressed \cite{GL}%
. They are also infinite, hence require renormalization. Note, however, that
since no extra divergencies are introduced by nuclear matter, the
vacuum-subtracted sum rules (\ref{GMOR0}-\ref{GMOR1}) are well-defined:

\begin{eqnarray}
&&-m_u\left( \langle \overline{u}u\rangle _C-\langle \overline{u}u\rangle _{%
{\rm vac}}\right) -m_d\left( \langle \overline{d}d\rangle _C-\langle 
\overline{d}d\rangle _{{\rm vac}}\right)  \nonumber \\
&=&\sum_{j^0}\left| \langle j^0|J_{5,0}^0(0)|C\rangle \right|
^2-\sum_{j^0}\left| \langle j^0|J_{5,0}^0(0)|{\rm vac}\rangle \right| ^2,
\label{GMOR0sub}
\end{eqnarray}
and similarly for the sum rule (\ref{GMOR1}).

\subsection{Additional sum rules}

\label{sec:add} Repeating the steps of the previous section on Eq. (\ref
{commV1}) we arrive at the sum rule

\begin{eqnarray}
&&-(m_u-m_d)\langle \overline{u}u-\overline{d}d\rangle _C-{\rm vac} 
\nonumber \\
&=&\sum_{j^{-}}{\rm sgn}(E_{j^{-}})\left| \langle j^{-}|J_0^{-}(0)|C\rangle
\right| ^2+\sum_{j^{+}}{\rm sgn}(E_{j^{+}})\left| \langle
j^{+}|J_0^{+}(0)|C\rangle \right| ^2-{\rm vac},  \label{add1}
\end{eqnarray}
where ${\rm vac}$ means the vacuum subtraction as in Eq. (\ref{GMOR0sub}).
Here the intermediate states have quantum numbers of the $\delta $ meson, ($%
I^G(J^{PC})=1^{-}(0^{++})$).

Subsequent sum rules are obtained from Eqs. (\ref{su2xsu2}). The derivation
repeats the steps of App. \ref{app:sr}. We obtain two sum rules involving
the isovector density, $\rho _{I=1}=\frac 12\langle u^{\dagger }u-d^{\dagger
}d\rangle _C$:

\begin{eqnarray}
\langle u^{\dagger }u-d^{\dagger }d\rangle _C &=&\sum_{j^{-}}\frac
1{|E_{j^{-}}|}\left| \langle j^{-}|J_{5,0}^{-}(0)|C\rangle \right|
^2-\sum_{j^{+}}\frac 1{|E_{j^{+}}|}\left| \langle
j^{+}|J_{5,0}^{+}(0)|C\rangle \right| ^2,  \label{vec1} \\
\langle u^{\dagger }u-d^{\dagger }d\rangle _C &=&\sum_{j^{-}}\frac
1{|E_{j^{-}}|}\left| \langle j^{-}|J_0^{-}(0)|C\rangle \right|
^2-\sum_{j^{+}}\frac 1{|E_{j^{+}}|}\left| \langle j^{+}|J_0^{+}(0)|C\rangle
\right| ^2.  \label{vec2}
\end{eqnarray}
Sum rules (\ref{vec1}) and (\ref{vec2}) involve excitations with the
quantum
numbers of $\pi $ and $\delta $, respectively. These sum rules require no
vacuum subtraction, since the left-hand sides involve the matrix element of
the conserved vector-isovector current ({\em \ }$\rho _{I=1}=\langle
C|J_0^0(0)|C\rangle $). If the state $|C\rangle $ is isosymmetric, {\em i.e. 
}$\rho _{I=1}=0$, then the above relations are trivial and just reflect the
isospin symmetry of the excitation spectrum.

\section{Formal results from the sum rules}

\label{sec:formal} For the discussion of this section it is convenient to
rewrite the sum rules using the identities 
\begin{eqnarray}
E_{j^a}\langle j^a|J_{5,0}^a(0)|C\rangle &=&\langle j^a|\left[ H_{{\rm QCD}%
},J_{5,0}^a(0)\right] |C\rangle =\langle j^a|\overline{\psi }(0)i\gamma
_5\left\{ \tau ^a/2,{\cal M}\right\} \psi (0)|C\rangle ,  \nonumber \\
E_{j^a}\langle j^a|J_0^a(0)|C\rangle &=&\langle j^a|\left[ H_{{\rm QCD}%
},J_0^a(0)\right] |C\rangle =\langle j^a|\overline{\psi }(0)\left[ \tau ^a/2,%
{\cal M}\right] \psi (0)|C\rangle .  \label{iden}
\end{eqnarray}
Then Eq. (\ref{GMOR0}) becomes 
\begin{equation}
-m_u\langle \overline{u}u\rangle _C-m_d\langle \overline{d}d\rangle _C-{\rm %
vac}=\sum_{j^0}\frac 1{E_{j^0}^2}\left| \langle j^0|\left( m_u\overline{u}%
i\gamma _5u-m_d\overline{d}i\gamma _5d\right) |C\rangle \right| ^2-{\rm vac},
\label{GMOR0r}
\end{equation}
where $j^0$ labels all excitations with the quantum numbers
of $\pi ^0$, Eqs. (%
\ref{GMOR1},\ref{vec1}) give 
\begin{eqnarray}
&&-\langle \overline{u}u+\overline{d}d\rangle _C-{\rm vac}  \label{GMOR1r} \\
&=&\sum_{j^{-}}\frac{m_u+m_d}{2|E_{j^{-}}|E_{j^{-}}}\left| \langle j^{-}|%
\overline{d}i\gamma _5u|C\rangle \right| ^2+\sum_{j^{+}}\frac{m_u+m_d}{%
2|E_{j^{+}}|E_{j+}}\left| \langle j^{+}|\overline{u}i\gamma _5d|C\rangle
\right| ^2-{\rm vac},  \nonumber \\
2\rho _{I=1} &=&\sum_{j^{-}}\frac{\left( m_u+m_d\right) ^2}{2|E_{j^{-}}|^3}%
\left| \langle j^{-}|\overline{d}i\gamma _5u|C\rangle \right| ^2-\sum_{j^{+}}%
\frac{\left( m_u+m_d\right) ^2}{2|E_{j^{+}}|^3}\left| \langle j^{+}|%
\overline{u}i\gamma _5d|C\rangle \right| ^2,  \label{vec1r} \\
&&  \nonumber
\end{eqnarray}
where $j^{\pm }$ label all excitations with the quantum numbers
of $\pi ^{\pm },$
and finally Eqs. (\ref{add1},\ref{vec2}) give
\begin{eqnarray}
&&-\langle \overline{u}u-\overline{d}d\rangle _C-{\rm vac}  \label{addr} \\
&=&\sum_{j^{-}}\frac{m_u-m_d}{2|E_{j^{-}}|E_{j^{-}}}\left| \langle j^{-}|%
\overline{d}u|C\rangle \right| ^2+\sum_{j^{+}}\frac{m_u-m_d}{%
2|E_{j^{+}}|E_{j+}}\left| \langle j^{+}|\overline{u}d|C\rangle \right| ^2-%
{\rm vac},  \nonumber \\
2\rho _{I=1} &=&\sum_{j^{-}}\frac{\left( m_u-m_d\right) ^2}{2|E_{j^{-}}|^3}%
\left| \langle j^{-}|\overline{d}u|C\rangle \right| ^2-\sum_{j^{+}}\frac{%
\left( m_u-m_d\right) ^2}{2|E_{j^{+}}|^3}\left| \langle j^{+}|\overline{u}%
d|C\rangle \right| ^2,  \label{vec2r}
\end{eqnarray}
where $j^{\pm }$ label all excitations with the quantum numbers
of $\delta ^{\pm}$.

We stress that the above sum rules are valid for all values of current quark
masses, {\em i.e}. not necessarily in the chiral ($m_u+m_d\rightarrow 0$) or
isovector ($m_u-m_d\rightarrow 0$) limits, and hold for {\em all} densities $%
\rho _B$ and $\rho _{I=1}$.

\subsection{Chiral limit at finite density in isospin-symmetric medium}

\label{sec:chir} Now we are going to explore several formal predictions
following from Eqs. (\ref{GMOR0r},\ref{GMOR1r},\ref{addr}). The method has
been discussed in detail in Refs. \cite{tdcwb:isovector,algnjl}. First, we
analyze the case when the state $|C\rangle $ carries no isovector density,
such as the vacuum or symmetric nuclear matter. To simplify notation we take
the strict isovector limit $m_u=m_d=\overline{m}$. In this case of exact
isospin symmetry of the Hamiltionian, as well as of
the state $|C\rangle $, the
excitation spectrum is invariant under isospin rotations, and clearly $%
E_{j^0}=E_{j^{+}}=E_{j^{-}}$. Also, $\langle \overline{u}u\rangle _C=\langle
\overline{d}d\rangle _C=\langle \overline{q}q\rangle _C$. Sum rule (\ref
{GMOR0r}) becomes
\begin{equation}
-2\langle \overline{q}q\rangle _C-{\rm vac}=\sum_{j^0}\frac{\overline{m}}{%
E_{j^0}^2}\left| \langle j^0|\left( \overline{u}i\gamma _5u-\overline{d}%
i\gamma _5d\right) |C\rangle \right| ^2-{\rm vac.}  \label{simGMOR}
\end{equation}
As long as the chiral symmetry is broken, $\langle \overline{q}q\rangle _C$
is non-zero in the chiral limit $\overline{m}\rightarrow 0$. As already
mentioned, the vacuum subtraction terms are of order $\overline{m}$, thus
are chirally small. Therefore, to match the chiral dimensions on both sides
of Eq. (\ref{simGMOR}), there must exist a state, denoted as $\pi ^0$, for
which $\frac{\overline{m}}{E_{\pi ^0}^2}\left| \langle \pi ^0|\left(
\overline{u}i\gamma _5u-\overline{d}i\gamma _5d\right) |C\rangle \right|
^2\sim 1$. Since the matrix element $\langle \pi ^0|\left( \overline{u}%
i\gamma _5u-\overline{d}i\gamma _5d\right) |C\rangle $ is finite in the
chiral limit, then it follows that $E_{\pi ^0}\sim \sqrt{\overline{m}}$.
Thus, we obtain the same chiral scaling as in the vacuum, where in the
chiral limit we have $m_\pi =\sqrt{2\overline{m}\langle \overline{q}q\rangle
_0}/F_\pi \sim \sqrt{\overline{m}}$. By isospin symmetry we have
\begin{equation}
E_{\pi ^0}=E_{\pi -}=E_{\pi +}\sim \sqrt{\overline{m}}.  \label{isoscal}
\end{equation}
Note that this result is true for any baryon density.

In principle, in the dense medium there could be more than one state
contributing to the sum rule (\ref{simGMOR}) in the chiral limit. It is
known that many-body effects of the Fermi sea can induce additional branches
of excitations, and we could have several states scaling as (\ref{isoscal}).
Whether or not this occurs is a complicated dynamical issue. The formal
result states that there exists at least one state scaling as (\ref{isoscal}%
) in the chiral limit.

\subsection{Chiral limit at finite density in isospin-asymmetric medium}

\label{sec:chir2} As shown in Ref. \cite{tdcwb:isovector}, in medium with
finite isovector density, $\rho _{I=1}\neq 0$, the behavior of charged
excitations in the chiral limit is radically different from the isosymmetric
case (\ref{isoscal}). First, an obvious remark is that since the medium
state $|C\rangle $ breaks the isospin invariance, the isospin symmetry of
excitations is broken. In fact, at low densities \cite{Lutz} one can relate
the splitting of $E_{\pi +}$ and $E_{\pi -}$ to the Weinberg-Tomosawa term
in the $\pi $-$N$ scattering, and obtain
\begin{equation}
E_{\pi +}-E_{\pi -}=\frac{\rho _{I=1}}{F_\pi ^2}.  \label{WT}
\end{equation}
In this approach one takes the low-density limit{\em \ prior }to the chiral
limit. Eq. (\ref{WT}) shows that for negative $\rho _{I=1}$ at small
densities we have $E_{\pi -}>E_{\pi +}$. However, Eq. (\ref{WT}) cannot be
used at large densities.

Now, following Ref. \cite{tdcwb:isovector}, we assume that the isospin
density is fixed, and employ sum rules (\ref{GMOR1r}-\ref{vec1r}). Without
loss of generality we can assume that $\rho _{I=1}<0$, as in the
case of neutron
stars or large nuclei. Since $\rho _{I=1}$ is an external property of the
system,{\em \ i.e.} independent of the chiral parameter, it is treated as
large (finite) in the chiral limit. Then, also the {\em isovector chemical
potential}, $\mu _{I=1}$, defined as the minimum energy needed to lower the
isospin by one unit, is finite in the chiral limit.\footnote{%
Although we have not been able to prove this statement from first
principles, one can present a number of physical arguments in its favor. In
the $\rho $-exchange model discussed in Ref. \cite{tdcwb:isovector}, when an
object of isospin $I_3$ is placed in the isospin-asymmetric medium, the
energy gain is equal to $g_\rho ^2/m_\rho ^2$ $\rho _{I=1}I_3$, and the
corresponding chemical potential is $\mu _{I=1}=g_\rho ^2/m_\rho ^2\,\rho
_{I=1}\simeq \rho _{I=1}/(2F_\pi ^2)$ , where the last equality follows from the
KSFR\ relation. This shows that finite $\rho _{I=1}$ in the chiral limit
implies finite $\mu _{I=1}$. Another example is provided by the Fermi-gas
model discussed in this paper. The expression for the chemical potential is
(see Sec. \ref{sec:njl} for notation) $%
\mu _{I=1}=\rho +\sqrt{k_u^2+M_u^2}-$ $\sqrt{k_d^2+M_d^2}$, and it is finite
when $k_u\neq k_d$, i.e. when $\rho _{I=1}$ is finite.} The excitation
energies of positive (negative) isospin can now be written as $E_{j^{\pm
}}=\mu _{I=1} \pm \delta E_{j^{\pm }}$, with $\delta E_{j^{\pm }}\geq 0$ by
definition of the chemical potential. Hence, for the medium with $\rho
_{I=1}<0$, or $\mu _{I=1}<0$, only the positive-isospin excitation energies
can vanish. The negative-isospin excitation energies cannot vanish,
including the case of the chiral limit. Hence we arrive at the result that
for $\rho _{I=1}<0$ the negative-charge excitations are chirally large,
\begin{equation}
E_{j-}\sim 1.  \label{neg}
\end{equation}
Since the matrix elements $\langle j^a|\overline{u}i\gamma _5d|C\rangle $
are non-singular in the chiral limit, the sum over negative-isospin
excitations, $j^{-}$, in Eqs. (\ref{GMOR1r}-\ref{vec1r}) does not contribute
as $m_u+$ $m_d\rightarrow 0$. Therefore in the chiral limit the
negative-charge states do not contribute at all to the sum rules (\ref
{GMOR1r}-\ref{vec1r}). For the sum rules to hold, there must exist
positive-charge states scaling appropriately in the chiral limit \cite
{tdcwb:isovector}. Assuming there is only one such state, labeled $\pi ^{+}$%
, we find that in the chiral limit

\begin{eqnarray}
-\langle \overline{u}u+\overline{d}d\rangle _C &=&\frac{m_u+m_d}{2|E_{\pi
^{+}}|E_{\pi +}}\left| \langle \pi ^{+}|\overline{u}i\gamma _5d|C\rangle
\right| ^2  \label{GMOR1z} \\
-2\rho _{I=1} &=&\frac{\left( m_u+m_d\right) ^2}{2|E_{\pi ^{+}}|^3}\left|
\langle \pi ^{+}|\overline{u}i\gamma _5d|C\rangle \right| ^2.  \label{vec1z}
\\
&&  \nonumber
\end{eqnarray}
Equations (\ref{GMOR1z}-\ref{vec1z}) give immediately

\begin{eqnarray}
E_{\pi +} &=&\frac{(m_u+m_d)\langle \overline{u}u+\overline{d}d\rangle _C}{%
2\rho _{I=1}}\sim {\cal O}(m_u+m_d),  \label{eplus} \\
\left| \langle \pi ^{+}|\overline{u}i\gamma _5d|C\rangle \right| ^2 &=&\frac{%
(m_u+m_d)\left| \langle \overline{u}u+\overline{d}d\rangle _C\right| ^3}{%
2\rho _{I=1}^2}\sim {\cal O}(m_u+m_d).  \label{matplus}
\end{eqnarray}
This is totally different from the ``usual'' behavior in the chiral limit,
Eq. (\ref{isoscal}): the excitation energy of the $\pi ^{+}$ mode scales as
the current quark mass itself, $m_u+m_d$, and not $\sqrt{m_u+m_d}$.

The formal case where more than one state contributes to the sum rules
(\ref{GMOR1r}-\ref{vec1r}) in
the chiral limit has been analyzed in Ref. \cite{tdcwb:isovector} in the
following way: Assume that the excitation energies of one of these modes
scales as $E_{\pi +}\sim \left( m_u+m_d\right) ^\alpha $, and the
corresponding matrix element scales as $\left| \langle \pi ^{+}|\overline{u}%
i\gamma _5d|C\rangle \right| \sim \left( m_u+m_d\right) ^\beta $. Since the
matrix element is not singular in the chiral limit, one has $\beta \geq 0$.
The RHS of the sum rule (\ref{vec1r}) contains, in the chiral limit, only the
positive-charge contributions, which are negative-definite. Therefore no
cancellations between contributions of various modes can occur in the chiral
limit. The requirement of matching chiral powers on both sides of
Eq. (\ref{vec1r})
gives $3\alpha =2+2\beta $, from which we conclude that $\alpha \geq 2/3$.
The positive-charge contributions to the RHS of the sum rule (\ref{GMOR1r})
may contain
both positive and negative terms, since the sign of $E_{j^{+}}$ is not
constrained. This means that there can be cancellations of the leading
chiral powers of the positive-charge modes contributing to the sum rule in
the chiral limit. Matching of the chiral powers on both sides of
Eq. (\ref{GMOR1r})
yields therefore the inequality $0\geq 1-2\alpha +2\beta $, which means,
together with the previously-derived equality $3\alpha =2+2\beta $, that $%
\alpha \leq 1$. In case of no cancellations $\alpha =1$. Combining the above
results one gets \cite{tdcwb:isovector}
\begin{equation}
2/3\leq \alpha \leq 1.  \label{ineq}
\end{equation}
Note that these inequalities are nontrivial, since in the vacuum the
corresponding power is $\alpha =1/2<2/3$.

Carrying a similar analysis for the neutral pionic excitation in
isospin-asymmetric medium we find that in the chiral limit it scales the
usual way according to Eq. (\ref{isoscal}), {\em i.e.} as in the
isosymmetric case. This is clear, since the medium does not break the third
component of isospin.

We conclude this section with several comments. Firstly, the behavior of
charged pionic modes in the chiral limit is radically different when the
medium breaks the isospin symmetry. For $\rho _{I=1}<0$ the negative-charge
excitation has finite energy in the chiral limit, Eq. (\ref{neg}), and the
positive-charge excitation becomes very soft, with excitation energy scaling
as Eq. (\ref{eplus}) (if there is a single mode contributing in the chiral
limit) or according to Eq. (\ref{ineq}) (if there are more modes).
Furthermore, as we will show in the model calculation in the following
sections, the nature of this soft mode can be quite complicated: it need not
be the excitation branch connected to the pion in the vacuum, but the
spin-isospin sound mode, resulting from collective effects in the Fermi sea.

\subsection{Isovector limit at finite density}

\label{sec:isovector} Now we pass to the analysis of Eq. (\ref{addr}-\ref
{vec2r}) in the isovector limit $m_u-m_d\rightarrow 0$. To our knowledge,
results of this section are novel. The method is the same as in the previous
sections, with the obvious difference that now we compare the powers of $%
m_u-m_d$ rather than $m_u+m_d$ on both sides of the sum rules (\ref{addr}-%
\ref{vec2r}).

In isosymmetric medium $\langle \overline{u}u-\overline{d}d\rangle _C\sim
m_u-m_d$. Thus, the powers of $m_u-m_d$ in Eq. (\ref{addr}) are matched with 
\begin{equation}
E_{\delta _{}^a}\sim {\cal O}(1),\;a=0,+,-.  \label{a0sym}
\end{equation}
This result is not surprising: since $\delta $ is not a Goldstone boson,
there is no reason for its mass to vanish in the vacuum, or in isosymmetric
medium.

In asymmetric medium the situation is different. Since $\rho _{I=1}=\langle
u^{\dagger }u-d^{\dagger }d\rangle _C$ is large in the isovector limit, {\em %
i.e.} does not vanish when $m_u-m_d\rightarrow 0$, also $\langle \overline{u}%
u-\overline{d}d\rangle _C$ is large. This is a direct effect of the
asymmetry of the Fermi sea, since the Fermi momenta of $u$ and $d$ quarks
(or protons and neutrons) are not equal. As a result, the Fermi seas of
particles with opposite isospin contribute differently to $\langle \overline{%
u}u\rangle _C$ and $\langle \overline{d}d\rangle _C$. An explicit example is
provided later on in Eq. (\ref{ubaru}).

Repeating the steps of Sec. \ref{sec:chir2} we now find that for $\rho
_{I=1}<0$ there must exist at least one state $\delta ^{+}$ whose excitation
energy scales as $m_u-m_d$ in the isovector limit. If there is only one such
state, then we find

\begin{eqnarray}
E_{\delta ^{+}} &=&\frac{(m_u-m_d)\langle \overline{u}u-\overline{d}d\rangle
_C}{2\rho _{I=1}}\sim {\cal O}(m_u-m_d),  \label{a0plus} \\
\left| \langle \delta ^{+}|\overline{u}d|C\rangle \right| ^2 &=&\frac{%
(m_u-m_d)\left| \langle \overline{u}u-\overline{d}d\rangle _C\right| ^3}{%
2\rho _{I=1}^2}\sim {\cal O}(m_u-m_d),  \label{mata0}
\end{eqnarray}
Equation (\ref{a0plus}) shows that there is an exact zero mode in the case
of the strict isovector limit $m_u=m_d$. Such collective modes have been
known to occur in many-body physics \cite{abrik,bashkin,provid}. If more
such states exist, then, in analogy to the case of pionic excitations ({\em %
cf. }the discussion above Eq. (\ref{ineq})), we find that the excitation
energies of these states scale as $\left( m_u-m_d\right) ^\alpha $, with $%
2/3\leq \alpha \leq 1$. This result is nontrivial, since in the vacuum the
corresponding power is $\alpha =0<2/3$.

\subsection{Comments}

\label{sec:arbitrarym}We end the formal part of this paper with several
comments. We stress again that the sum rules of Sec. \ref{sec:formal} are
valid for arbitrary values of the current quark masses, not necessarily in
the chiral or isovector limit, and for arbitrary densities. All kinds of
intermediate states contribute to the sum rules: quasiparticles (poles),
which can come in multiple branches, 2p-2h continuum,{\em \ etc. }

Another remark concerns the sign of the excitation energy of a mode. As
already noticed in Refs. \cite{tdcwb:isovector,algnjl}, the charged
excitation may have negative excitation energy. Note, however, that this
does not mean that the system is unstable. This is because charged
excitations change the isospin of the system. Suppose we request the state $%
|C\rangle $ to be the ground state of matter with isospin {\em constrained}
to the value $I_3$. A charged excitation in the sum rules involves isospin $%
I_3\pm 1$. Thus, its isospin is outside the constrained value, and even if
the energy of the mode is lower than the energy of $|C\rangle $, it does not
mean instability of $|C\rangle $. As a corollary, we notice that states of
negative excitation energy cannot be the soft modes of Eq. (\ref{eplus})
in the chiral
limit. We conclude this from Eq. (\ref{eplus}). Since the quark condensate
is negative, we get (for media with negative isospin density) that
$E_{\pi^+}>0$.

\section{The Nambu--Jona-Lasinio model}

\label{sec:njl} In the remaining parts of this paper we are going to
illustrate in detail the general results discussed above by a model
calculation. We will consider quark matter in the Nambu--Jona-Lasinio model 
\cite{njl}. This model acquired great popularity in recent years as a
framework for calculations of meson and baryon properties, also in the
nuclear medium \cite
{Lutz,Hatsuda85,BMZ:temp,ReinDang,Vogl1,Lutz92,Jaminon2,RuivoHiller,Klevansky}
described as a Fermi gas of quarks. Although the description of the nuclear
medium by a Fermi gas of quarks is certainly not realistic (unless,
perhaps, at very high densities where nucleons deconfine), the model is
well-suited for our theoretical purpose: it is consistent with the
constraints of chiral symmetry. Therefore it complies to the current-algebra
relations, and the sum rules of Sec. \ref{sec:formal} hold.

The Lagrangian of the $SU(2)\otimes SU(2)$ Nambu--Jona-Lasinio model is
\begin{eqnarray}
{\cal L} &=&\bar q({\rm i}\gamma ^\mu \partial _\mu -{\cal M})q+{\frac{%
G_\sigma }2}\left( (\bar qq)^2+(\bar q{\rm i}\gamma _5\tau ^aq)^2\right) +%
\frac{{G}_\delta }2\left( (\bar q\tau ^aq)^2+(\bar q{\rm i}\gamma
_5q)^2\right)  \nonumber \\
&&\ \ \ -{\frac{G_\rho }2}\left( (\bar q\gamma _\mu \tau ^aq)^2+(\bar
q\gamma _5\gamma _\mu \tau ^aq)^2\right) -{\frac{G_\omega }2}(\bar q\gamma
_\mu q)^2\;,  \label{Lagsu2}
\end{eqnarray}
where $q$ is the quark field, ${\cal M}$ is the quark mass matrix, and $G$'s
denote the coupling constants in various channels. Using the usual technique
of the Hartree approximation one arrives at self-consistency equations for
the values of the scalar-isoscalar field $S$, the neutral component of the
scalar-isovector field $\delta $, the time component of the neutral
vector-isovector field $\rho $, and the time component of the
vector-isoscalar field $\omega $:
\begin{eqnarray}
S &=&\frac{m_u+m_d}2-G_\sigma \langle \overline{u}u+\bar dd\rangle ,\;\delta
=\frac{m_u-m_d}2-G_\delta \langle \overline{u}u-\bar dd\rangle ,\;  \nonumber
\\
\rho &=&2G_\rho \langle u^{+}u-d^{+}d\rangle ,\;\omega =G_\omega \langle
u^{+}u+d^{+}d\rangle .  \label{meanf}
\end{eqnarray}

For the numerical study in the examples below we use the following two
parameter sets, fixed by meson properties in the vacuum: \newcounter{aux} %
\setcounter{aux}{\arabic{equation}} \renewcommand{\theequation}{%
\Roman{equation}} \setcounter{equation}{0}
\begin{eqnarray}
G_\sigma &=&7.55{\rm GeV}^{-2},\;G_\delta =5.41{\rm GeV}^{-2},\;G_\rho =7.09%
{\rm GeV}^{-2},\;  \nonumber \\
\;\Lambda &=&750{\rm MeV},\;m_u=2.52{\rm MeV},\;m_d=4.52{\rm MeV},
\label{setI} \\
&&  \nonumber \\
G_\sigma &=&4.35{\rm GeV}^{-2},\;G_\delta =3.34{\rm GeV}^{-2},\;G_\rho =12.4%
{\rm GeV}^{-2},\;  \nonumber \\
\Lambda &=&954{\rm MeV},\;m_u=1.03{\rm MeV},\;m_d=3.03{\rm MeV}.
\label{setII}
\end{eqnarray}
\setcounter{equation}{\arabic{aux}} \renewcommand{\theequation}{%
\arabic{equation}} The value of $G_\omega $ and the $\omega $ field are not
relevant, since we will look for excitations carrying no baryon number.
Parameter set \ref{setI} has been used in Ref. \cite{Klimt} to fit the
mesonic properties: $m_\pi $, $F_\pi $, $m_\rho =765{\rm MeV}$,\footnote{%
Note that in this fit the $\rho $ meson lies just above the $q\bar q$
production threshold, hence the fit is somewhat problematic. However, this
issue is not of much relevance for our illustrative application of the model.%
}, and $m_\eta =519{\rm MeV}$. This fits four parameters out of original
six. The remaining two parameters are chosen in such a way that $S=361{\rm %
MeV}$, and the current quark masses $m_d$ and $m_u$ are arbitrarily split by 
$2{\rm MeV}$. Parameter set \ref{setII} also fits $m_\pi $, $F_\pi $, $%
m_\eta $ and $S=361{\rm MeV}$, but not $m_\rho $. It uses a much larger
value for $G_\rho $. Such larger values are needed if
one wishes to fit the $a_{11}$
$\pi $-$\pi $ scattering length \cite{a11}. Following Ref. \cite{Klimt}, we
regularize the model using the sharp 3-momentum cut-off. Our results do not
qualitatively depend on the choice of the regulator, since the constraints
of current algebra are satisfied. The 3-momentum cut-off obeys these
requirements, in particular it leads to correct Ward identities \cite{Lutz}.

The scalar and vector densities of the $u$ and $d$ quarks are equal to 
\begin{eqnarray}
\langle \overline{u}u\rangle &=&2N_c\int \frac{d^3k}{(2\pi )^3}\frac{M_u}{%
\sqrt{k^2+M_u^2}}\left( \Theta (k_u-|k|)-\Theta (\Lambda -|k|)\right) ,\; 
\nonumber  \label{eq:dens0} \\
\;\;\;\langle \overline{d}d\rangle &=&2N_c\int \frac{d^3k}{(2\pi )^3}\frac{%
M_d}{\sqrt{k^2+M_d^2}}\left( \Theta (k_d-|k|)-\Theta (\Lambda -|k|)\right)
,\;  \nonumber  \label{ubaru} \\
\langle u^{+}u\rangle &=&2N_c\int \frac{d^3k}{(2\pi )^3}\Theta
(k_u-|k|)\;,\quad \langle d^{+}d\rangle =2N_c\int \frac{d^3k}{(2\pi )^3}%
\Theta (k_d-|k|)\;,  \label{udagu}
\end{eqnarray}
where $\Lambda $ is the sharp three-momentum cut-off, and $k_u$ and $k_d$
are the $u$ and $d$ quark Fermi momenta, and $\Theta $ is the step function.
We have introduced scalar self-energies of $u$ and $d$ quarks, given by 
\begin{equation}
M_u=S+\delta ,\qquad M_d=S-\delta .  \label{masses}
\end{equation}
Self-consistency requires that the quark propagators be evaluated with
mean-fields (\ref{meanf}): 
\begin{equation}
S_{u/d}^{-1}=p-\gamma _0(\pm \frac \rho 2+\omega )-M_{u/d}+i\varepsilon \,%
{\rm sgn}(\mu _{u/d}-p_0)\;,  \label{eq:S}
\end{equation}
where $\mu _u$ and $\mu _d$ are the chemical potentials of the $u$ and $d$
quarks.

\section{Mean fields in medium}

\label{sec:field} We introduce the $x$ and $y$ variables, 
\begin{equation}
x=\frac{\rho _B}{\rho _0}=\frac{\rho _u+\rho _d}{N_c\rho _0},\quad y=\frac{%
\rho _d}{\rho _u+\rho _d},  \label{xy}
\end{equation}
where $\rho _0=0.17{\rm fm}^{-3}$ is the nuclear saturation density, $\rho
_B $ is the baryon number density, and $\rho _{u/d}$ are the quark
densities. The variable $y$ measures the isospin asymmetry of the medium. In
symmetric medium $y=$ $\frac 12$, and in pure neutron matter $\rho _d=2\rho
_u$ and $y=\frac 23$. The isospin density of the system can be written as $%
\rho _{{\rm I=1}}=\frac 12(\rho _u-\rho _d)=N_c\rho _0x(\frac 12-y)$. In our
study we fix the $x$ and $y$ variables, hence we examine properties of quark
matter at a given baryon density and isospin asymmetry.

The first task is to find the mean fields by solving Eqs. (\ref{meanf}). For
the field $\rho $, which couples to the isospin current, we get immediately $%
\rho =4G_\rho \rho _{{\rm I=1}}=4G_\rho N_c\rho _0x(\frac 12-y)$. The values
of $S$ and $\delta $ are determined by solving numerically the first two of
equations (\ref{meanf}). Results for $M_u$, $M_d$ and $-\frac 12\rho $ are
displayed in Fig. 1. 
For isosymmetric matter (top row) $M_u$ is practically equal to $M_d$, and
the small splitting is caused by the current quark mass difference, $%
m_d-m_u=2{\rm MeV}$. For $y=\frac 23$ (middle row) and $x$ in the range $1$
to $4$ we find that $M_u$ is greater than $M_d$ by $10-20{\rm MeV}$. At
maximum asymmetry (bottom row) the $u$ quark is heavier than the $d$ quark
by $100-200{\rm MeV}$. There is a simple physical argument why $M_u>M_d$ at $%
y$ above $\frac 12$: the $d$ quarks are more abundant, and it is
energetically preferable for the system to make them lighter.\footnote{%
Note that although the values of constituent quark masses in the two lower
rows of Fig. 1 
decrease with density, in the strict sense
it does not mean chiral restoration. This is because
chiral symmetry cannot be restored when isospin is broken. Indeed, if $Q_3
|C\rangle \ne 0$, then by charge algebra $[Q^+_5,Q^-_5]|C\rangle \ne 0$,
hence we cannot restore chiral symmetry, in which case we would have $%
Q^a_5|C\rangle = 0$, a=1,2,3.}

\begin{figure}[tbp]
\vspace{0mm} \epsfxsize = 12.5 cm \centerline{\epsfbox{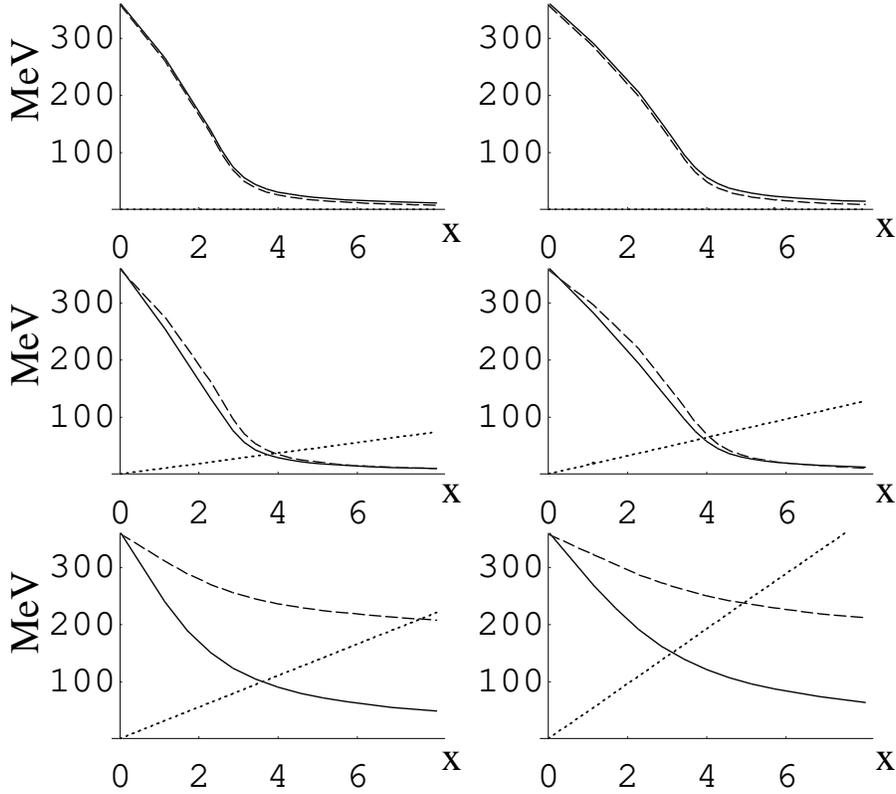}} %
\vspace{0mm} \label{fig:meanf}
\caption{The quark scalar self-energies, $M_d$ (solid line) and $M_u$
(dashed line), and the vector self-energy of the $d$ quark, $-\frac 12\rho $
(dotted line) plotted as functions of $x=\rho _B/\rho _0$ for several values
of $y$. The top plots are for $y=1/2$ (isosymmetric matter), the middle
plots are for $y=2/3$ (pure neutron matter), and the bottom plots are for $%
y=1$ (pure $d$-quark matter). The plots on the left (right) are for
parameter set I (II).}
\end{figure}
We note that the field $S$ has a large value of the order ${\cal O}(1)$ if
the chiral symmetry is broken. Otherwise it is of the order ${\cal O}%
(m_u+m_d).$ The field $\delta $ is large only if $\langle \overline{u}u-\bar
dd\rangle \sim {\cal O}(1)$, which occurs in isospin-asymmetric medium. In
isosymmetric medium $\langle \overline{u}u-\bar dd\rangle \sim {\cal O}%
(m_u-m_d)$, and $\delta $ is small, of the order ${\cal O}(m_u-m_d)$.

\section{Meson propagators in medium}

\label{sec:mesons} As explained in Appendix \ref{app:sr}, only excitations
``at rest'' enter the sum rules. Furthermore, we shall only consider the
charged meson propagators, since the interesting effects take place for that
case.

In the case of no vector-isovector interactions ({\em i.e.} $G_\rho =0$),
the one-quark-loop inverse pion propagator acquires a simple form $%
1-G_\sigma J_{\pi \pi }$, where 
\begin{equation}
J_{\pi \pi }(q)=-i{\rm Tr}\int \frac{d^4k}{(2\pi )^4}\gamma _5S_u(k+\frac
12q)\gamma _5S_d(k-\frac 12q).  \label{jpp0}
\end{equation}
In presence of vector-isovector interactions there is a complication due to
the well-known mechanism of mixing of $\pi $ and the longitudinal component
of the $A_1$ meson. In that case in order to find excitation energies one
has to find {\em zeros} of the determinant of the inverse $\pi -A_1$
propagator matrix, $D_\pi $ (see {\em e.g.} Ref.~\cite{Lutz} for details
concerning this problem). The explicit form of the determinant is given in
Eq.(\ref{eq:detpi}).

\begin{figure}[tbp]
\vspace{0mm} \epsfxsize = 8 cm \centerline{\epsfbox{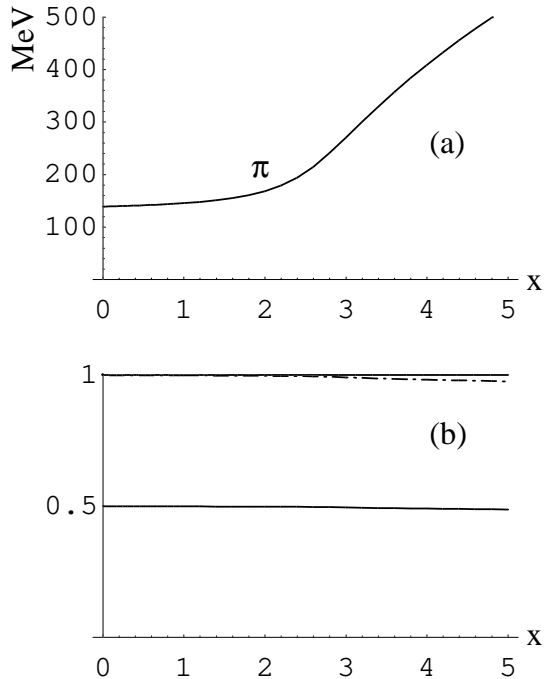}} \vspace{0mm} 
\label{fig:pi}
\caption{Charged pion in isosymmetric matter ($y=1/2$), parameter set I. (a)
Charged pion excitation energy plotted as fuction of $x$. (b) The
contribution of the pion poles to sum rule (\ref{GMOR1}): the solid line
shows the contribution of $\pi^{+}$ (which equals to the contribution of $%
\pi^{-}$), the dash-dotted line shows the combined contribution of the two
poles.}
\end{figure}
It is worthwhile to look at the analytic structure of $D_\pi $, or
equivalently, $J_{\pi \pi }$, in the variable $q_0$. The matter state $%
|C\rangle $ consists of the Fermi seas of $d$ and $u$ quarks, with $k_d>k_u$%
, as well as of the Dirac sea occupied down to the cut-off $\Lambda $. A
positive-charge Fermi sea excitation moves a quark from the occupied $d$
level to an unoccupied $u$ level. Pauli blocking allows this when 
\begin{equation}
\rho +\sqrt{k_d^2+M_u^2}-\sqrt{k_d^2+M_d^2}<q_0<\rho +\sqrt{k_u^2+M_u^2}-%
\sqrt{k_u^2+M_d^2}.  \label{fcut}
\end{equation}
Thus, within these boundaries $D_\pi (q_0)$ possesses a cut. The cuts
associated with the Dirac sea are within the boundaries: 
\begin{eqnarray}
\rho -\sqrt{\Lambda _{}^2+M_u^2}-\sqrt{\Lambda _{}^2+M_d^2} &<&q_0<\rho -%
\sqrt{k_d^2+M_u^2}-\sqrt{k_d^2+M_d^2},  \nonumber \\
\rho +\sqrt{k_u^2+M_u^2}+\sqrt{k_u^2+M_d^2} &<&q_0<\rho +\sqrt{\Lambda
_{}^2+M_u^2}+\sqrt{\Lambda _{}^2+M_d^2}.  \label{dirc}
\end{eqnarray}

In the delta channel we proceed analogously. We define 
\begin{equation}
J_{\delta \delta }(q)=-i{\rm Tr}\int \frac{d^4k}{(2\pi )^4}S_u(k+\frac
12q)S_d(k-\frac 12q).  \label{jdd0}
\end{equation}
For the case $G_\rho =0$ the inverse charged $\delta $-meson propagator is $%
1-G_\delta J_{\delta \delta }$. For finite $G_\rho $ there occurs mixing
between the $\delta $ meson and the longitudinal component of the $\rho $
meson. This mixing is proportional to the mean field $\delta ,$ hence it is
small, of the order of ${\cal O}(m_u-m_d)$ in isosymmetric medium. The
stated behavior can be promptly seen from Eq. (\ref{wardvector2}). If the
medium is asymmetric, then the mean field $\delta $ is large, and such is
the $\delta -\rho $ mixing. The explicit form of the appropriate
determinant, $D_\delta $, is given in Eq.(\ref{eq:detdel}). The location of
the cuts of $D_\delta $ is of course the same as in the pion case.

\section{Mesons in symmetric matter}

\label{sec:symm} Figure 1 
shows the results of the numerical calculation of the charged pion
excitation in
\begin{figure}[tbp]
\vspace{0mm} \epsfxsize = 8 cm \centerline{\epsfbox{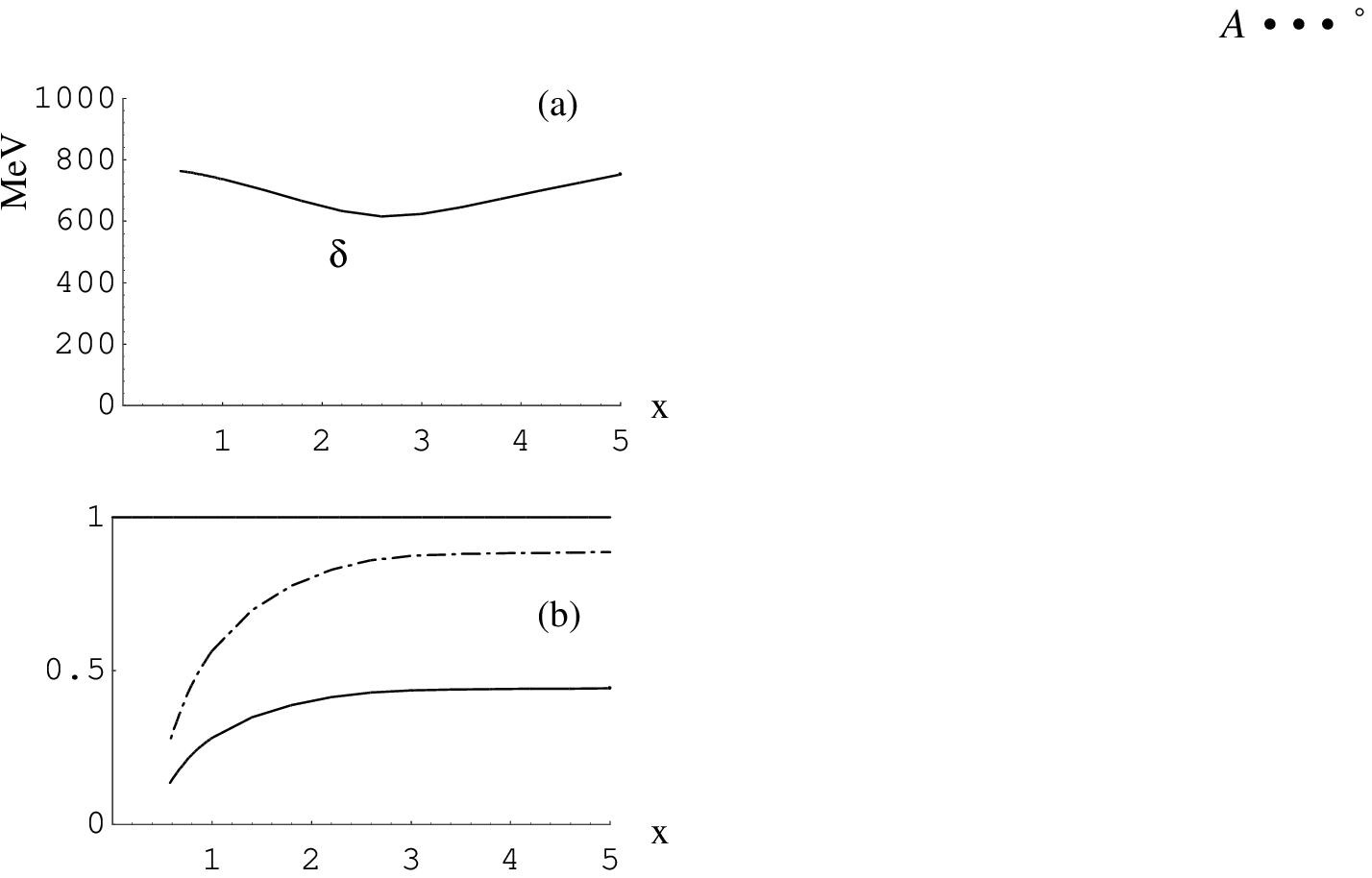}} \vspace{0mm}
\label{fig:delI12}
\caption{Charged $\delta$ meson in isosymmetric matter ($y=1/2$). (a) The
excitation energy plotted as fuction of $x$. (b) The contribution of the $%
\delta $ poles to sum rule (\ref{add1}): the solid line shows the
contribution of $\delta ^{+}$ (which equals to the contribution of $\delta^-$%
, the dash-dotted line shows the combined contribution of the two poles.}
\end{figure}
symmetric matter. In Fig. 2(a) 
we show the position of the charged pion excitation at rest (usually called
the in-medium pion mass) as a function of baryon density. The behavior is
the expected one \cite{Vogl1}, with the pion mass increasing slowly with the
baryon density up to about $x\simeq 2$. Above this point chiral symmetry is
restored,{\em \ i.e.} $S\sim {\cal O}(m_u+m_d)$, ({\em cf}. the upper left
Fig. 1%
) and the pion mass grows more rapidly.

Figure 2(b)
shows the anatomy of the in-medium GMOR sum rule
(\ref{GMOR1}). We note that up to $x=2$ practically all of the sum rule is
saturated by the charged pion poles. At larger $x$ some small ( a few per
cent) strength is carried by the cuts ({\em cf. } Eqs. (\ref{fcut}-\ref{dirc}%
)). We have verified for all other cases shown in this paper that the sum of
all pole and cut contributions to the sum rules adds up to 100\%. This serves
as a check of the numerical calculations.

The case of the charged $\delta $ excitation is displayed in Fig. 3.
This excitation emerges as a bound state from the $q\bar q$ continuum at $%
x\simeq 0.6$. Its mass decreases with the baryon density up to $x\simeq 2.5$%
, and then starts growing (Fig. 3%
(a)). The contributions to the sum rule (\ref{addr}) are shown in Fig. 3(b).
We can see, especially at lower values of $x$, that the pole
contribution fall short of saturating the sum rule. Continuum contributions
carry about 50\% at $x=1$ and about 15\% at $x>3$.

\section{Mesons in asymmetric matter}

\label{sec:asymm} In this section we come to the central part of our paper.
We will show that in our model the sum 
\begin{figure}[tbp]
\vspace{0mm} \epsfxsize = 8 cm \centerline{\epsfbox{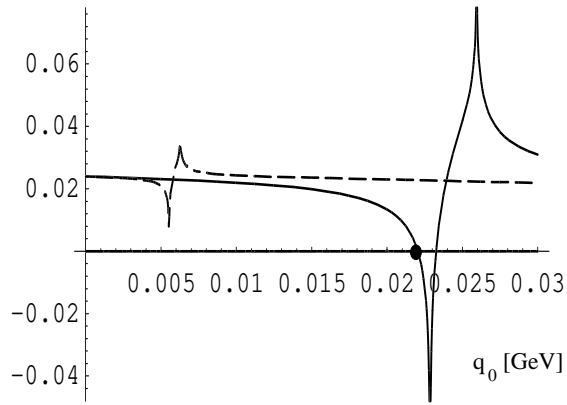}} \vspace{0mm}
\label{fig:depGd}
\caption{The real part of the pion determinant for $G_\delta =4.7{\rm GeV}%
^{-2}$ (dashed line) and $G_\delta =9.4{\rm GeV}^{-2}$ (solid line), plotted
as a function of $q^0$ for $y=\frac 23$. Other parameters are: $G_\sigma
=11.7{\rm GeV}^{-2}$, $\Lambda =619{\rm MeV}$, and $m_u=m_d=5.7{\rm MeV}$.
For large values of $G_\delta $ a zero of the inverse pion proparator is
induced in the vicinity of the $\overline{d}u$, denoted by a blob. This
indicates the presence of the ``spin-isospin sound''.}
\end{figure}
rules from Sec. \ref{sec:formal} are, for the case of isospin asymmetric
medium, satisfied in a non-trivial way. This involves a collective state,
specific for asymmetric medium. As explained {\em e.g.} in \cite
{migdal,rhowil} in the framework of conventional nuclear physics, it is
possible for the pion propagator in neutron matter to have an additional
pole at very low excitation energies. Such an excitation is known as the 
{\em spin-isospin sound}. We will show that this phenomenon occurs in our
model.

The existence of collective modes in our model is related to the presence of
the Fermi sea cut (\ref{fcut}). Figure 4 
shows the {\em real} part of the determinant $D_\pi $ for $y=\frac 23$ (pure
neutron matter) and two sample parameter choices, plotted as a function of
the energy variable $q^0$ in the region of the Fermi sea cut. Let us first
look at the solid line, corresponding to parameters with a large coupling
constant $G_\delta $. The presence of the cut manifests itself by the two
cusps. The imaginary part of $D_\pi $ is nonzero in the region between the
two cusps, and vanishes outside. We notice that a zero of $D_\pi $ exists in
the vicinity of the cut, indicated in the figure by a blob. This zero, at $%
q^0=23{\rm MeV}$, corresponds to the energy of the spin-isospin sound mode.
The dashed line, corresponding to lower $G_\delta $, also has cusps, but no
zero of $D_\pi $ exists. This can be understood as follows: the cut region
is wider and the function at the cusps acquires higher and lower values as the
splitting of the scalar self-energies $M_u$ and $M_d$ is larger ({\em cf.}
Eq.(\ref{fcut})). This splitting is proportional to the mean field $\delta $%
, which increases with $G_\delta $, and with asymmetry $y$. Thus we have a
critical behavior: above some critical values of asymmetry $y$ and coupling $%
G_\delta $ the spin-isospin mode emerges. We denote it by $\pi _S$. In the
example shown in Fig. 4
the excitation energy of $\pi _S$ is lower than the left boundary of the
cut. We find this is the case for small values of the vector-isovector
coupling constant $G_\rho $. At sufficiently large values of $G_\rho $ the
collective mode emerges at energies larger than the right boundary of the
Fermi cut. In any case, the collective state lies very close to the Fermi
sea cut, with excitation energy of the order of $10{\rm MeV}$.

In addition to the collective mode $\pi _S$, there exist the usual two
charged pion branches, $\pi ^{+}$ and $\pi ^{-}$, with excitation energies
of the order of $m_\pi $. These branches connect to the vacuum pion as the
baryon density is lowered. Thus, depending on parameters and the value of $y,
$ we have, in our model, $2$ or $3$ branches of the charged pion excitations.

For the charged $\delta $ channel the situation is similar: for appropriate
parameters and $y>\frac 12$, a collective mode $\delta _S$ appears in
addition to the usual $\delta ^{+}$ and $\delta ^{-}$ modes.

\section{Sum rules in asymmetric medium}

\label{sec:srasymm} In this section we show the results of our numerical
study. For the case of pionic excitations these results have been already
reported in Ref. \cite{algnjl} (for the slightly different parameter cases
with $m_u=m_d$ ). 
\begin{figure}[tbp]
\vspace{0mm} \epsfxsize = 8 cm \centerline{\epsfbox{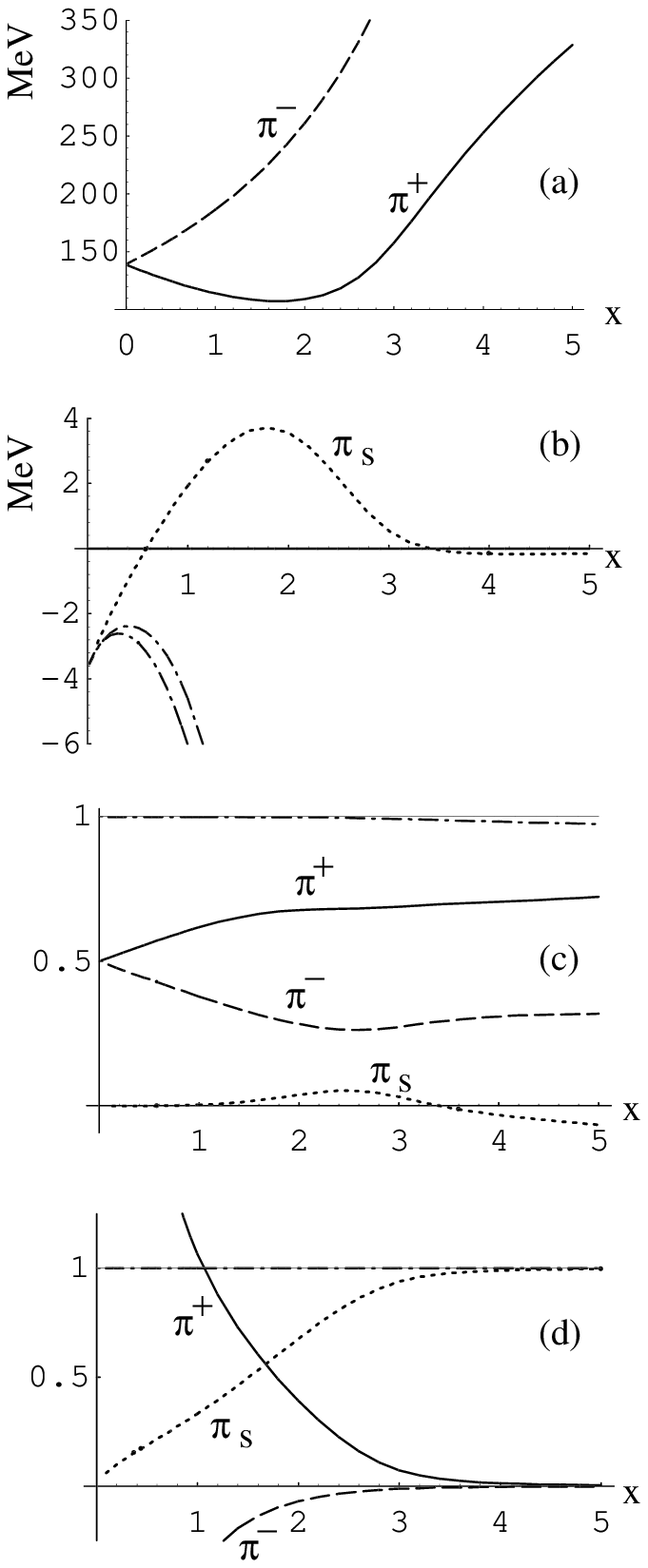}} \vspace{0mm} 
\label{fig:piI23}
\caption{Properties of charged pion excitations for $y=\frac 23$ and
parameter set \ref{setI}, plotted as a function of baryon density.
(a)~Excitation energies of $\pi ^{+}$ and $\pi ^{-}$. (b)~Excitation energy
of $\pi _S$ (dotted line) and the boundaries of the Fermi sea cut
(dot-dashed line). (c) The relative contribution of $\pi ^{+}$, $\pi ^{-}$
and $\pi _S$ to the in-medium GMOR sum rule (\ref{GMOR1}), and the total
contribution from the three poles (dot-dashed line). (d) The relative
contribution of $\pi ^{+}$, $\pi ^{-}$ and $\pi _S$ to the sum rule (\ref
{vec1}). The total contribution from the three poles (dot-dashed line)
practically saturates the sum rule.}
\end{figure}
Figure 5 
shows the results for the $\pi $ channel for the parameter set \ref{setI}.
Figure 5%
(a) shows the excitation energies of the usual branches, $\pi ^{+}$ and $\pi
^{-}$, and Fig. 5 
(b) shows the excitation energy of the collective $\pi _S$ mode. The
dashed-dotted lines show the boundaries of the Fermi-sea cut, (\ref{fcut}).
The collective mode emerges from the cut at a low value of the baryon
density. Its excitation energy is positive for $x$ between $0.6$ and $3.4$,
and negative otherwise. In Fig. 5%
(c) we show the relative contributions from the poles to the in-medium GMOR
sum rule, Eq. (\ref{GMOR1}), and the total contribution from the three
poles, indicated by the dash-dotted line. The poles practically saturate the
sum rule, leaving 1-2\% for the cuts at large values of $x$. The
contribution of $\pi _S$ to the sum rule (\ref{GMOR1}) is of the order of a
few per cent. Its sign follows the sign of the excitation energy in Fig. 5%
(c), as is apparent from Eq. (\ref{GMOR1}). Figure 5%
(d) shows the relative contribution of the poles to the sum rule (\ref{vec1}%
), and the total pole contribution, indicated by the dash-dotted line. We
note that this sum rule is saturated by the pole at the 99.9\% level -- the
cut contributions turn out to be very small. At larger value of $x$ the
collective $\pi _S$ mode dominates over the other modes, and for $x>3$ it
practically saturates the sum rule. We note that the sign of the
contributions is associated with the charge of the excitation, as is clear
from Eq. (\ref{vec1}).

\begin{figure}[tbp]
\vspace{0mm} \epsfxsize = 8 cm \centerline{\epsfbox{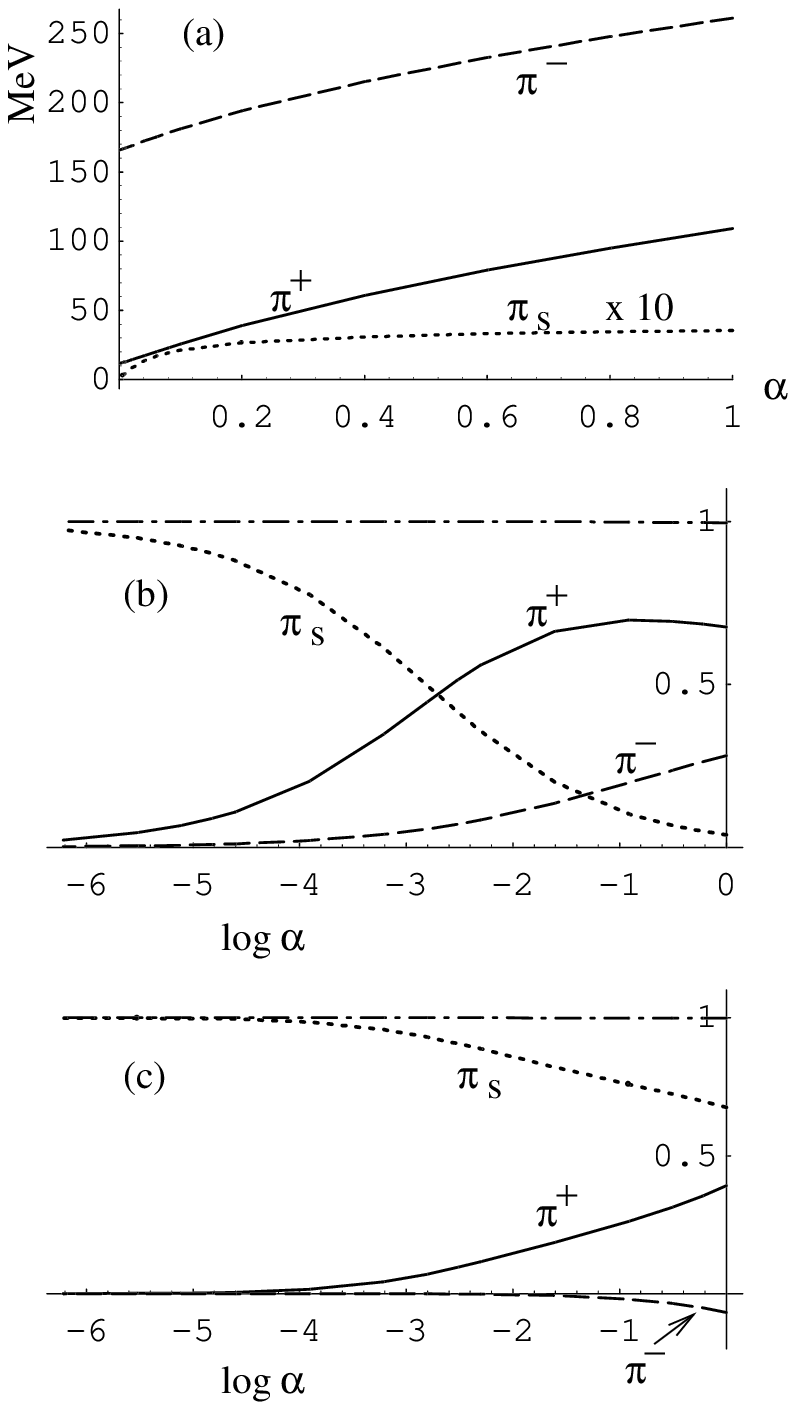}} \vspace{0mm}
\label{fig:piI23m}
\caption{Study of the chiral limit for $x=2$ and $y=\frac{2}{3}$. Convention
for line the same as in Fig.~5. (a) Excitation energies, (b) relative
contributions to the sum rule (\ref{GMOR1}), and (c) relative contributions
to the sum rule (\ref{vec1}), plotted functions of $\alpha$, Eq.~(\ref{alpha}%
). The spin-isospin sound mode $\pi_S$ is the chiral soft mode
of Eq. (\ref{eplus}).
Please, note that $\log \alpha$ means $\ln \alpha$.}
\end{figure}

Figure 6 
shows the result of a formal study of the chiral limit, $m_u+m_d\rightarrow 0
$. For fixed values of $y=\frac 23$ and $x=2$ we lower the value of 
\begin{equation}
\alpha =\frac{m_u+m_d}{m_u^{{\rm phys}}+m_d^{{\rm phys}}},  \label{alpha}
\end{equation}
\begin{figure}[tbp]
\vspace{0mm} \epsfxsize = 8 cm \centerline{\epsfbox{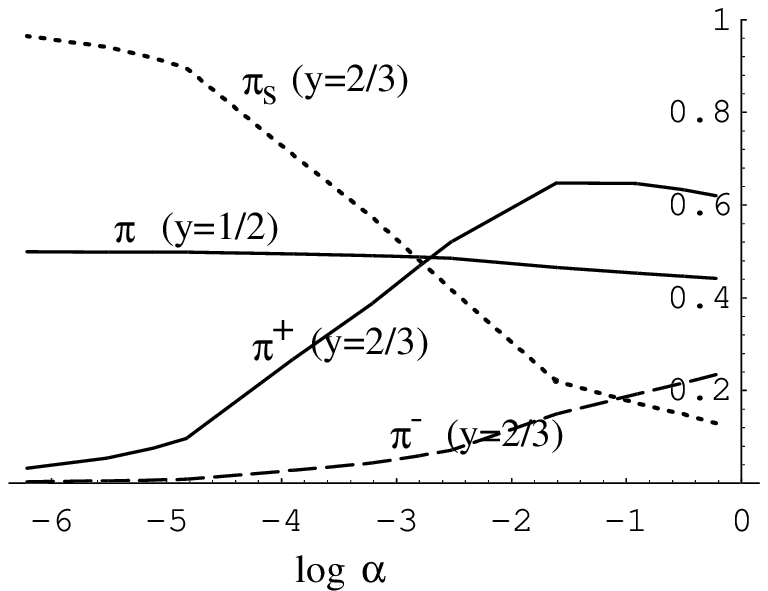}} \vspace{0mm} 
\label{fig:pidim}
\caption{Chiral dimensions of pionic modes plotted as functions of $\log
\alpha$, {\em cf.} Eq.~(\ref{alpha}). For the symmetric matter (middle solid
line, labeled $y=\frac{1}{2}$, the chiral dimension of the pionic
excitations tends to $\frac{1}{2}$ in the chiral limit ($\log \alpha \to -
\infty$). For asymmetric matter ($y=\frac{2}{3}$) the chiral dimension of $%
\pi_S$ tends to $1$, and the chiral dimensions of $\pi^+$ and $\pi^-$ tend
to $0$ in the chiral limit. Please, note that $\log \alpha$ means $\ln \alpha
$.}
\end{figure}
where here the superscript $^{{\rm phys}}$ denotes the values from the
parameter set (\ref{setI}). Figure 6%
(a) shows that as the value of $\alpha $ is decreased, the excitation of the
modes go down. The excitation energies of $\pi ^{+}$ and $\pi ^{-}$ modes go
to finite values at $\alpha \rightarrow 0$, and the excitation energy of $%
\pi _S$ goes to $0$. Hence $\pi _S$ is the chiral soft mode of Eq. (%
\ref{eplus}). Figures 6%
(b-c)show that in the chiral limit of $\alpha \rightarrow 0$ the $\pi _S$
saturates the sum rules (\ref{GMOR1},\ref{vec1}). However, this happens at
very low values of $\alpha $, around $\ln \alpha =-4$ or $-5$. Such values
of $\alpha $ would correspond to the vacuum value of the pion mass of the
order of $15{\rm MeV}$. This indicates that from the point of view of the
sum rules we are quite far away from the chiral limit with the physical
values of current quark masses,{\em \ i.e.} with $\alpha =1$.

\begin{figure}[tbp]
\vspace{0mm} \epsfxsize = 8 cm \centerline{\epsfbox{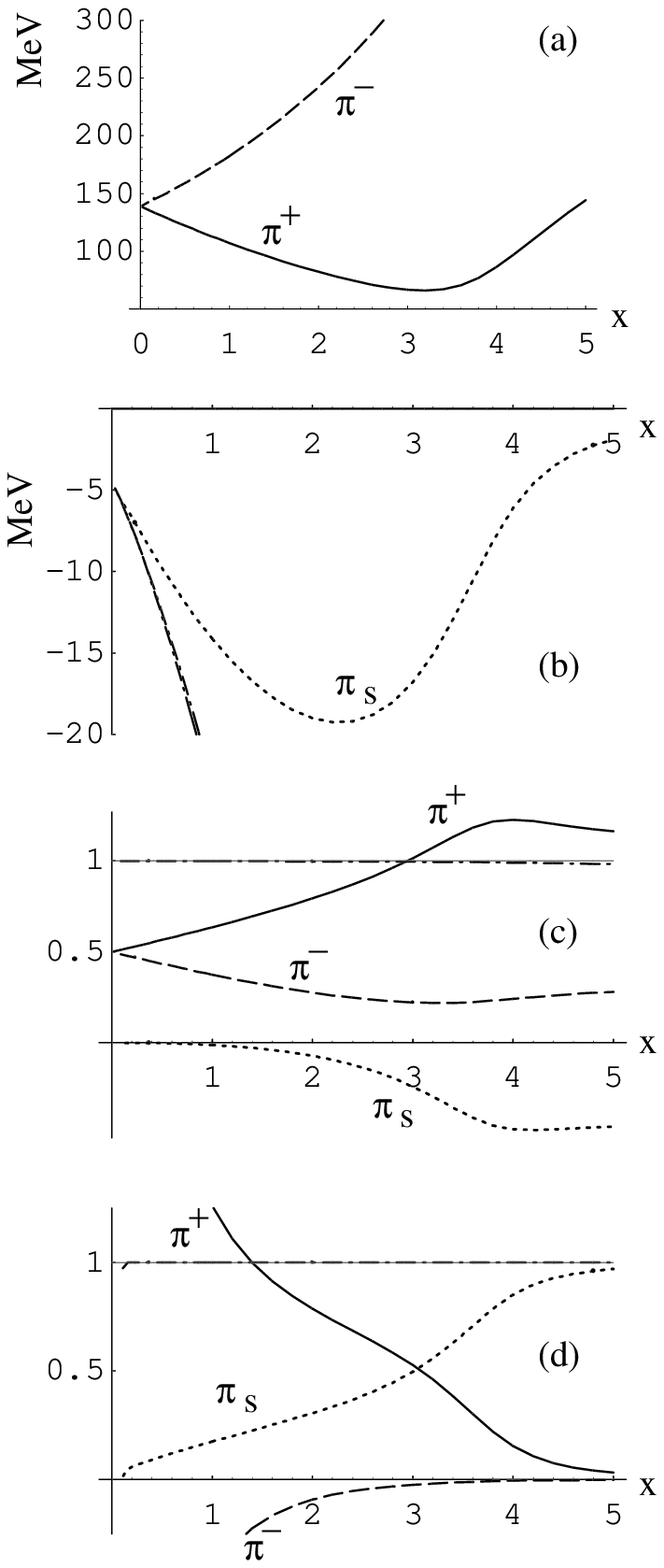}} \vspace{0mm}
\label{fig:piII23}
\caption{Same as Fig.~5 for parameter set \ref{setII}}
\end{figure}

In order to better illustrate this point we show in Fig. 7 
how the excitation energies of various modes approach the chiral limit.
Following Ref. \cite{tdcwb:isovector} let us introduce 
\mbox{${\rm dim}(X) = \lim_{\overline{m} \to
0}(\log{X}/\log{\overline{m}})$}, which we call the chiral dimension of
quantity $X$. In the chiral limit a quantity $X$ has some scaling with a
power of $m_u+m_d$. The function ${\rm dim}(X)$ extracts this power (for
instance in the vacuum ${\rm dim}(m_\pi )=1/2$). The dotted line in Fig. 7 
shows the chiral dimension of the excitation energy of $\pi _S$, which tends
to $1$ in the chiral limit, according to Eq. (\ref{eplus}). The chiral
dimensions of $\pi ^{+}$ and $\pi ^{-}$ go to $0$ in the chiral limit. The
solid line in the middle of the plot is for $\pi ^{+}$ or $\pi ^{-}$ in
symmetric matter, $y=\frac 12$. In that case, according to Eq. (\ref{isoscal}%
), the chiral dimension goes to $\frac 12$ in the chiral limit. 
\begin{figure}[tbp]
\vspace{0mm} \epsfxsize = 8 cm \centerline{\epsfbox{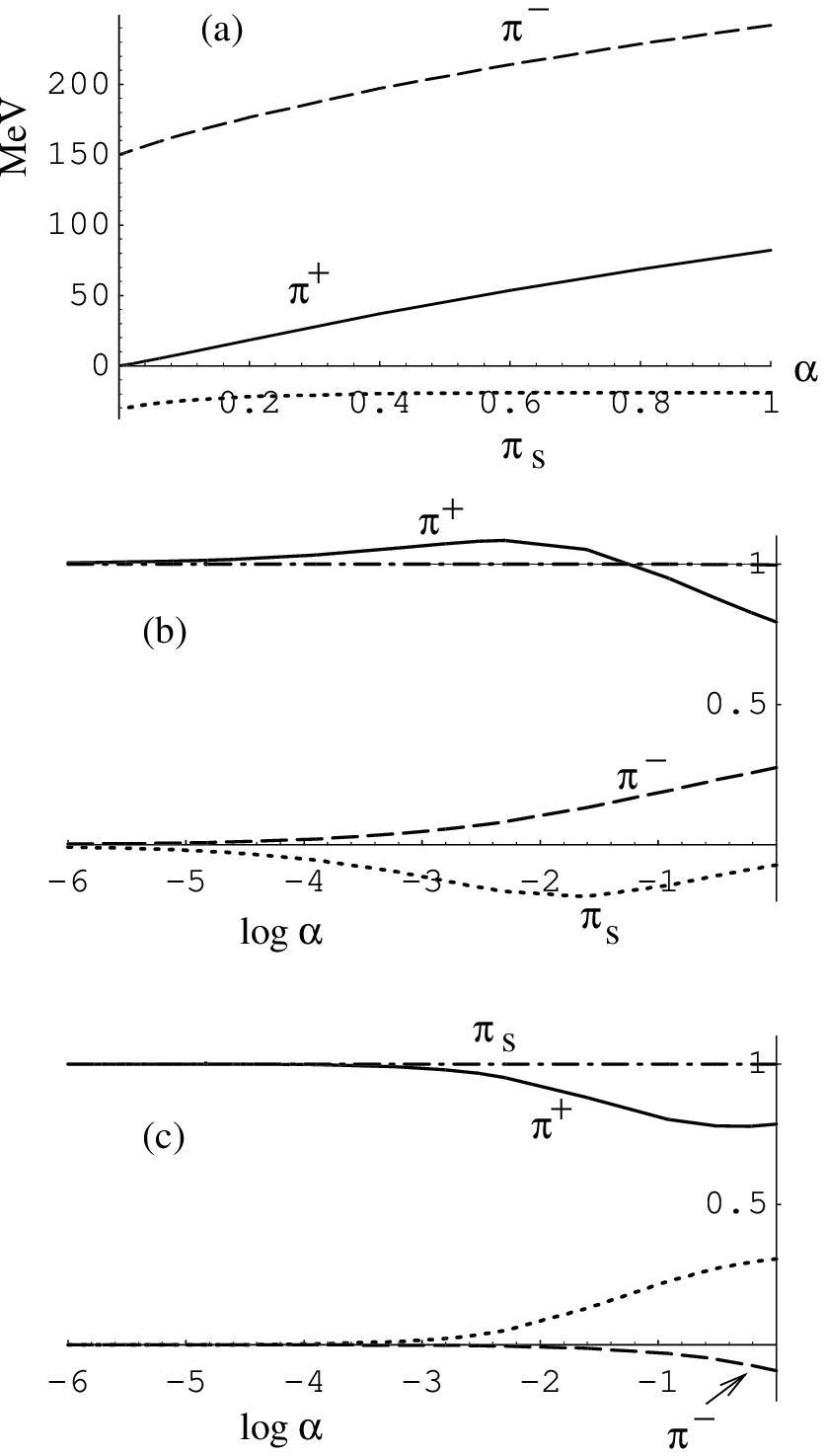}} %
\vspace{0mm} \label{fig:piII23m}
\caption{Same as Fig. 6 for parameter set \ref{setII}. The $\pi ^{+}$ mode
is the chiral soft mode of Eq. (\ref{eplus}).}
\end{figure}

Figures 8-9 
show the results for the parameter set \ref{setII}, also for $y=\frac 23$ as
in the case discussed above. The qualitative difference between the present
and the former case is that now the spin-isospin sound mode $\pi _S$ has
negative excitation energy for all values of $x$. Therefore in the chiral
limit it is the $\pi ^{+}$ mode, not $\pi _S$, which becomes the chiral
soft mode of Eq. (\ref{eplus}) (see Fig. 9 
). 
\begin{figure}[tbp]
\vspace{0mm} \epsfxsize = 8 cm \centerline{\epsfbox{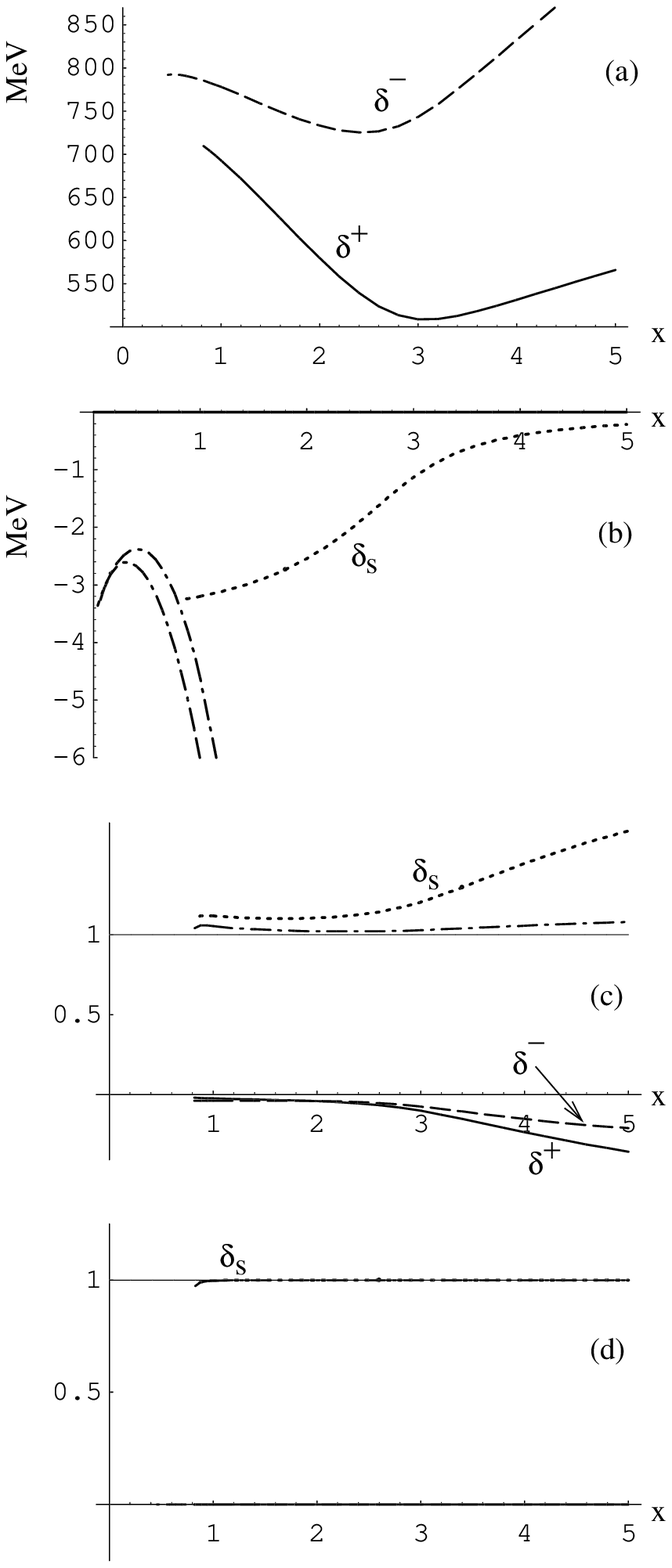}} \vspace{0mm}
\label{fig:delI23}
\caption{Properties of charged $\delta$ excitations for $y=\frac{2}{3}$ and
parameter set \ref{setI}, plotted as a function of baryon density.
(a)~Excitation energies of $\delta^+$ and $\delta^-$. (b)~Excitation energy
of $\delta_S$ (dotted line) and the boundaries of the Fermi sea cut
(dot-dashed line). (c) The relative contribution of $\delta^+$, $\delta^-$
and $\delta_S$ to the sum rule (\ref{add1}), and the total contribution from
the three poles (dot-dashed line). (d) The relative contribution of $%
\delta^+ $, $\delta^-$ (indistinguishable from $0$) and $\delta_S$ to the
sum rule (\ref{vec2}). The contribution of $\delta_S$ saturates the sum
rule. }
\end{figure}

Now we pass to the discussion of the $\delta $ channel, which is done for
the parameter set \ref{setI} only, and for $y=\frac 23$ . Figure 10 
(a) shows the excitation energies of the $\delta ^{+}$ and $\delta ^{-}$
branches. After emerging from the $q\bar q$ continuum their energies first
decrease until $x\sim 3$, and then start increasing. The collective mode $%
\delta _S$ emerges from the cut at $x\sim 0.75$ (Fig.
11(b)).
Its excitation energy is negative and small, less than $3{\rm MeV}$. Figures
10%
(c-d)
show the relative contributions to the sum rules (\ref{add1}) and (\ref{vec2}%
). We note that the $\delta _S$ mode plays a major role in sum rule (\ref
{add1}), and completely dominates sum rule (\ref{vec2}).

\begin{figure}[tbp]
\vspace{0mm} \epsfxsize = 9 cm \centerline{\epsfbox{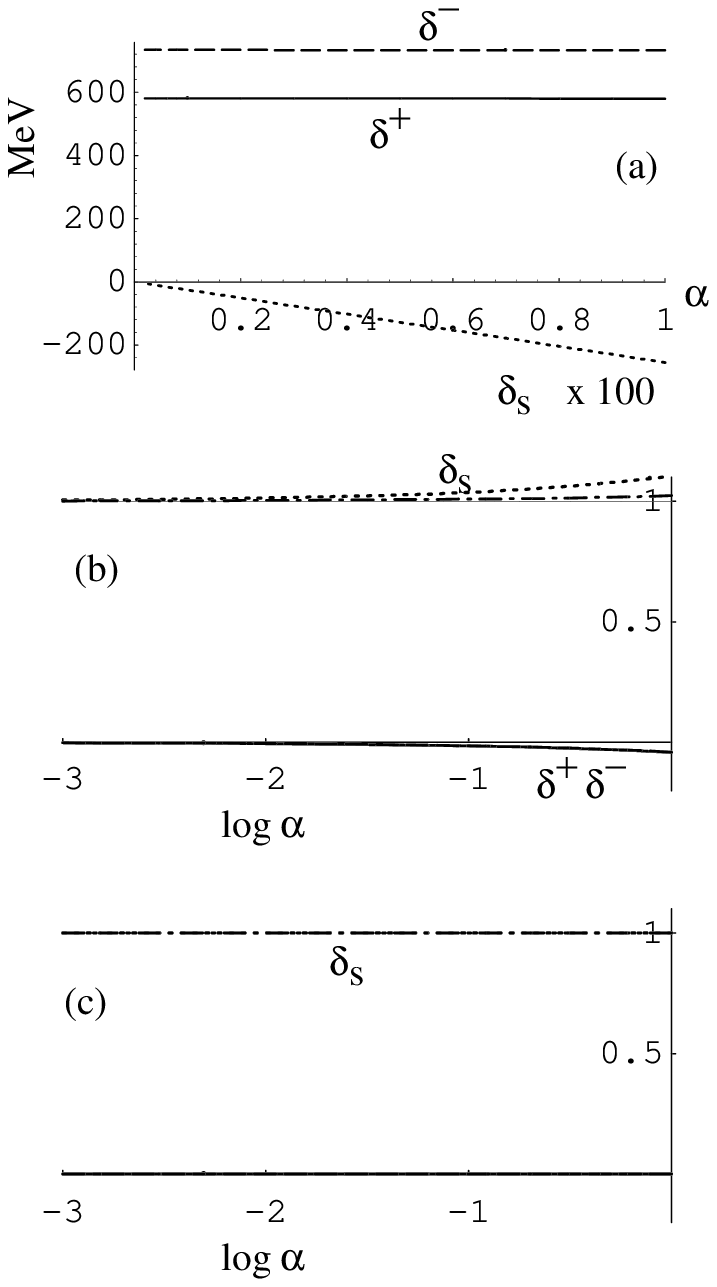}} %
\vspace{0mm} \label{fig:delI23m}
\caption{Study of the strict isovector limit for $x=2$ and $y=\frac{2}{3}$.
Convention for lines is the same as in Fig.~10. (a) Excitation energies, (b)
relative contributions to the sum rule (\ref{add1}), and (c) relative
contributions to the sum rule (\ref{vec2}), plotted as functions of $\alpha$%
, Eq.~(\ref{alpha2}). The spin-isospin sound mode $\delta_S$ is the
isovector soft mode of Eq. (\ref{a0plus}). It completely satisfies the sum rules in the
strict isovector limit. Please, note that $\log \alpha$ means $\ln \alpha$.}
\end{figure}

Figure 11 
shows the isovector limit for the $\delta $ channel. In this case 
\begin{equation}
\alpha =\frac{m_u-m_d}{m_u^{{\rm phys}}-m_d^{{\rm phys}}}.  \label{alpha2}
\end{equation}
We can see that the $\delta _S$ mode is the isovector soft mode of Eq.
(\ref{a0plus}). Its excitation energy drops linearly to $0$ as $\alpha $ is
decreased (Fig. 11%
(a)), and the sum rules are completely saturated by the $\delta _S$ mode in
the isovector limit of $\alpha \rightarrow 0$.

\section{Concluding remarks}

There are several messages which follow from our calculation.
Firstly, we note that in order to satisfy the current-algebraic sum
rules it is necessary to include all modes, in particular the spin-isospin
sounds. Certainly, a nuclear system is a very complicated object, and even
our simple model, treated at the 1p-1h level, has revealed a rich structure
of the excitation spectrum.
The power of the current-algebraic sum rules relies in the fact that
they relate in a non-trivial way
the properties of these excitations to the quark
condensate and the isospin density.

One may ask the following general questions: How far are we in a nuclear
system from the strict chiral limit ( $m_u+m_d\rightarrow 0$) and the strict
isovector limit ( $m_u-m_d\rightarrow 0$) in the real world, {\em i.e.} in a
dense nuclear system, and with the physical values of $m_u$ and $m_d$. The
results shown in Figs. 7 and 11
indicate, that
in moderately-dense isospin-asymmetric systems we are far away from the
chiral limit, and very close to the isovector limit. From Fig. 7
we find that the $\pi _S$ mode excitation energy scales linearly with $%
m_u+m_d$ starting from
$m_\pi \simeq \sqrt{e^{-6}}139{\rm MeV}\sim 7{\rm MeV}$,
much lower than the physical value. On the other hand, Fig.
10(a) shows that the excitation energy of the $\delta _S$ mode scales linearly
with $m_u-m_d$ already at physical value, corresponding to $\alpha =0$.

Another comment is relevant for application of effective chiral Lagrangians
to nuclear systems. In this approach one basically assumes that there is one
pion quasiparticle in a nuclear medium, albeit with modified properties
compared to the vacuum. In our model we find additional branches. Since they
contribute largely to the sum rules, they cannot be neglected. In an
effective model they should be included as additional degrees of freedom.

The final remark concerns strangeness. Although in this paper we have worked
for simplicity with two flavors, extension to three flavors is
straightforward. In fact, one can make a simple ``translation'' of the sum
rules of Sec. \ref{sec:formal} to the case of any flavor. For example,
changing the $d$ (or $u$) quark to $s$ we obtain the case of charged
(neutral) kaons. This is simply the replacement of $I$-spin by $U$ or $V$
spins. Note that nuclear matter is asymmetric with respect to $U$ and $V$
spins, therefore kaonic excitations on top of nuclear matter are parallel to
the case of charged pionic excitation on top of isospin-asymmetric matter.
We note that recently Refs. \cite{Ruivo} discussed the kaonic
excitations in the Fermi gas of quarks in the Nambu--Jona-Lasinio model.

\appendix

\section{Derivation of sum rules in medium}

\label{app:sr} In this appendix we explain the derivation of sum rules (\ref
{GMOR0}-\ref{vec2}). Although the technique is very well known, we believe
it is worthwhile to remind it in some greater detail in order to point out
the differences between the derivation in the vacuum and in a medium. The
first step in deriving the sum rules is to sandwich both sides of Eqs. (\ref
{commA0}-\ref{commV1}) by the medium state $|C\rangle $. On the RHS this
leads to a ``known'' quantity involving in-medium condensates $\langle C|%
\overline{u}u(0)|C\rangle \equiv \langle \overline{u}u\rangle _C$ and $%
\langle \overline{d}d\rangle _C$. Next, one inserts a
complete set of intermediate states $|\vec k,j\rangle $
between the current operators on the LHS.
These states are eigenstates of the momentum operator, $\widehat{P}$, and of
the Hamiltonian $H=\int d^3x{\cal H}(\vec x)$. They can be labeled by
additional quantum numbers, {\em e.g.} isospin. The medium state is also an
eigenstate of $\widehat{P}$ and $H$: 
\begin{equation}
\widehat{P}|C\rangle =\vec p_C|C\rangle ,\qquad H|C\rangle ={\cal E}_C|\vec
k,j\rangle .  \label{ph}
\end{equation}
For matter of a large volume $V$ the quantities $\vec p_C$ and ${\cal E}_C$
are proportional to $V$. It is convenient to measure the momentum and the
energy of intermediate states {\em relative} to the state $|C\rangle $, {\em %
\ i.e. } 
\begin{equation}
\widehat{P}|\vec k,j\rangle =\left( \vec k+\vec p_C\right) |\vec k,j\rangle
,\qquad H|\vec k,j\rangle =\left( E_j(\vec k)+E_C\right) |\vec k,j\rangle .
\label{ph2}
\end{equation}
Quantities $\vec k$ and $E_j(\vec k)$ form a Lorentz four--vector. Thus the
Lorentz-invariant mesure of integration is $\frac{d^3k}{|2E_j(\vec k)|}$,
and the unit operator can be decomposed as follows \cite{tdcwb:isovector}%
\footnote{%
The measure of integration is the same as for example in the case of phonon
excitations on top of a solid.}: 
\begin{equation}
1=\sum_j\int \frac{d^3k}{\left( 2\pi \right) ^32|E_j(\vec k)|}|\vec
k,j\rangle \langle \vec k,j|.  \label{unitop}
\end{equation}
We illustrate the method on Eq. (\ref{commV1}). We rewrite the LHS, insert
the unit operator (\ref{unitop}), express the charges by time-components of
currents, shift the coordinates of the currents with the translation
operator, and use Eq. (\ref{ph2}):
\begin{eqnarray*}
&&[Q^{-},[Q^{+},{\cal H}_{{\rm QCD}}(0)]]=\sum_j\int \frac{d^3k}{\left( 2\pi
\right) ^32|E_j(\vec k)|}\times  \\
&&\left( \langle C|Q^{+}|\vec k,j\rangle \langle \vec k,j|[Q^{-},{\cal H}%
(0)]|C\rangle -\langle C|[Q^{-},{\cal H}(0)]|\vec k,j\rangle \langle \vec
k,j|Q^{+}|C\rangle \right) = \\
&=&\sum_j\int \frac{d^3k}{\left( 2\pi \right) ^32|E_j(\vec k)|}\int d^3y\int
d^3x\times  \\
&&\left( \langle C|J_0^{+}(\vec y)|\vec k,j\rangle \langle \vec
k,j|[J_0^{-}(\vec x),{\cal H}(0)]|C\rangle -\langle C|[J_0^{-}(\vec y),{\cal %
H}(0)]|\vec k,j\rangle \langle \vec k,j|J_0^{+}(\vec x)|C\rangle \right) = \\
&=&\sum_j\int \frac{d^3k}{\left( 2\pi \right) ^32|E_j(\vec k)|}\int d^3y\int
d^3x\times  \\
&&\left( \langle C|e^{i\vec p_C\cdot \vec y}J_0^{+}(0)e^{-i\left( \vec
k+\vec p_C\right) \cdot \vec y}|\vec k,j\rangle \langle \vec k,j|e^{i\left(
\vec k+\vec p_C\right) \cdot \vec x}[J_0^{-}(0),{\cal H}(0)]e^{-i\vec
p_C\cdot \vec x}|C\rangle \right.  \\
&&\left. -\langle C|e^{i\vec p_C\cdot \vec y}[J_0^{-}(0),{\cal H}%
(0)]e^{-i\left( \vec k+\vec p_C\right) \cdot \vec y}|\vec k,j\rangle \langle
\vec k,j|e^{i\left( \vec k+\vec p_C\right) \cdot \vec x}J_0^{+}(0)e^{-i\vec
p_C\cdot \vec x}|C\rangle \right) = \\
&=&\sum_j\int \frac{d^3k}{2|E_j(\vec k=0)|}\delta ^3(\vec k)\left( \langle
C|J_0^{+}(0)|\vec k=0,j\rangle \langle \vec k=0,j|[J_0^{-}(0)(0),\int
\!\!d^3x{\cal H}(0)]|C\rangle \right.  \\
&&-\left. \langle C|[J_0^{-}(0)(0),\int \!\!d^3x{\cal H}(0)]|\vec
k=0,j\rangle \langle \vec k=0,j|J_0^{+}(0)|C\rangle \right) = \\
&=&\sum_j\frac 1{2|E_j(\vec k=0)|}\left( E_j(\vec k=0)\langle
C|J_0^{+}(0)|\vec k=0,j\rangle \langle \vec k=0,j|J_0^{-}(0)|C\rangle
\right.  \\
&&+\left. E_j(\vec k=0)\langle C|J_0^{-}(0)|\vec k=0,j\rangle \langle \vec
k=0,j|J_0^{+}(0)|C\rangle \right) = \\
&=&\sum_{j^{+}}\frac 12{\rm sgn}E_{j^{-}}\left| \langle
j^{-}|J_0^{-}(0)|C\rangle \right| ^2+\sum_{j^{+}}\frac 12{\rm sgn}%
E_{j^{+}}\left| \langle j^{+}|J_0^{+}(0)|C\rangle \right| ^2. \\
&&
\end{eqnarray*}
In the last line we have decomposed the sum over indices $j$ into the sum
over positive and negative isospin excitations. We have introduced the
short-hand notation $|j^{+}\rangle $ and $|j^{-}\rangle $ for such
excitations with relative momentum $\vec k=0$, and denoted their excitation
energies by $E_{j^{+}}$ and $E_{j^{-}}$. This completes the derivation of
the sum rule (\ref{add1}). With all other sum rules the steps are exactly
the same as described above.

\section{Charged meson propagators in medium}

\label{app:mes} The quark bubble for a meson channel is defined as 
\begin{equation}
J_{\Gamma \Gamma ^{\prime }}(q)=-i{\rm Tr}\int \frac{d^4k}{(2\pi )^4}\Gamma
S_u(k+\frac 12q)\overline{\Gamma }^{\prime }S_d(k-\frac 12q),  \label{bub}
\end{equation}
where $\overline{\Gamma }=\gamma _0\Gamma ^{\dagger }\gamma _0$ \cite{Klimt}%
. We use $\Gamma =\gamma _5$ in the pion vertex, which allows to get rid of
factors of $i$ in Ward identities below. For the considered case of $\vec q=0
$ the $\pi $-$A_1$ mixing involves the time components of the axial
propagator, $J_{A_1A_1}^{00}$, and the mixed propagator, $J_{\pi A_1}^0$.
The determinant of the inverse $\pi $-$A_1$ propagator is equal to
\begin{equation}
D_\pi (q^0)=(1-G_\sigma J_{\pi \pi }(q^0))(1+G_\rho
J_{A_1A_1}^{00}(q^0))+G_\sigma G_\rho (J_{\pi A_1}^0(q^0))^2.
\label{detpigen}
\end{equation}
The signs follow the convention for signs of the coupling constants in Eq. (%
\ref{Lagsu2}). The following Ward identities hold among the bubble functions
\cite{Klimt}:
\begin{eqnarray}
(q_0-\rho )J_{A_1A_1}^{00}(q^0) &=&2SJ_{\pi A_1}^0(q^0)+2(u^{\dagger
}u-d^{\dagger }d), \nonumber \\
(q_0-\rho )J_{\pi A_1}^0(q^0) &=&2SJ_{\pi \pi }^{}(q^0)+2(\bar uu+\bar dd),
\label{wardaxial}
\end{eqnarray}
where $\rho $ and $S$ are defined in Eq.(\ref{meanf}). These identities
follow from the general requirements of chiral symmetry \cite{Klimt}. They
can be explicitly verified to hold with our choice of the 3-momentum
regulator. Using Eqs. (\ref{wardaxial},\ref{meanf}) we can rewrite Eq. (\ref
{detpigen}) as
\begin{eqnarray}
D_\pi (q_0) &=&\frac{q_0}{q_0-\rho }\left\{ \frac{m_u+m_d}{2S}+\left( \frac{%
2(m_u+m_d)SG_\sigma ^{-1}G_\rho }{q_0(q_0-\rho )}-1\right) \times \right. 
\nonumber  \label{eq:detpi0} \\
&&\left. \left[ G_\sigma J_{\pi \pi }(q_0)-\frac{2S-(m_u+m_d)}{2S}\right]
\right\} \;.  \label{eq:detpi}
\end{eqnarray}
This form is convenient, since it involves only one bubble function, $J_{\pi
\pi }$, which has the explicit form
\begin{eqnarray}
J_{\pi \pi }(q_0) &=&4N_c\int_{k_u}^\Lambda \frac{d^3k}{(2\pi )^3}\frac{%
(\rho -q_0)+2\delta M_u/\sqrt{k^2+M_u^2}}{(\rho -q_0)^2+2(\rho -q_0)\sqrt{%
k^2+M_u^2}+4S\delta }  \nonumber \\
\ &+(u\to &d,\;\;\rho \to -\rho ,\;\;\delta \to -\delta ,\;\;q_0\to -q_0)\;.
\label{eq:jpp}
\end{eqnarray}
The zeros of $D_\pi (q_0)$ correspond to poles of the mixed charged $\pi $-$%
A_1$ propagator. The pole contributions to sum rules (\ref{GMOR1},\ref{vec1}%
) are explicitly given by the expression
\begin{eqnarray}
\!\!\!\!\!\!\!\!\!{\rm sgn}(E_{j^{\pm }})\left| \langle j^{\pm }\mid J_{5,0}^{\pm }\mid
C\rangle \right| ^2 &=&-\left. \frac{m_u+m_d}{q_0-\rho }\frac{\left[ SJ_{\pi
\pi }(q_0)-G_\sigma ^{-1}(S-(m_u+m_d)/2)\right] }{dD_\pi (q_0)/dq_0}\right|
_{q_0=E_j^{\pm }}\;.  \nonumber \\
&&
\label{eq:expoleax}
\end{eqnarray}

In the $\delta $-$\rho $ channel we obtain, if full analogy to Eq. (\ref
{detpigen}-\ref{eq:expoleax}),
\begin{equation}
D_\delta (q^0)=(1-G_\delta J_{\delta \delta }(q^0))(1+G_\rho J_{\rho \rho
}^{00}(q^0))+G_\delta G_\rho (J_{\delta \rho }^0(q^0))^2.  \label{detdelgen}
\end{equation}
Through the use of Ward identities
\begin{eqnarray}
(q_0-\rho )J_{\rho \rho }^{00}(q^0) &=&2\delta J_{\delta \rho
}^0(q^0)+2(u^{\dagger }u-d^{\dagger }d), \nonumber \\
(q_0-\rho )J_{\delta \rho }^0(q^0) &=&2\delta J_{\delta \delta
}^{}(q^0)+2(\bar uu-\bar dd),  \label{wardvector2}
\end{eqnarray}
where $\rho $ and $\delta $ are defined in Eq.(\ref{meanf}), we can rewrite
Eq. (\ref{detdelgen}) as
\begin{eqnarray}
D_\delta (q_0) &=&\frac{q_0}{q_0-\rho }\left\{ \frac{m_u-m_d}{2\delta }%
+\left( \frac{2(m_u-m_d)\delta G_\delta ^{-1}G_\rho }{q_0(q_0-\rho )}%
-1\right) \times \right.  \nonumber  \\
&&\left. \left[ G_\delta J_{\delta \delta }(q_0)-\frac{2\delta -(m_u-m_d)}{%
2\delta }\right] \right\} \;,  \label{eq:detdel}
\end{eqnarray}
where $J_{\delta \delta }$ is explicitly given by
\begin{eqnarray}
J_{\delta \delta }(q_0) &=&4N_c\int_{k_u}^\Lambda \frac{d^3k}{(2\pi )^3}%
\frac{(\rho -q_0)+2SM_u/\sqrt{k^2+M_u^2}}{(\rho -q_0)^2+2(\rho -q_0)\sqrt{%
k^2+M_u^2}+4S\delta }  \nonumber \\
\ &+(u\to &d,\;\;\rho \to -\rho ,\;\;\delta \to -\delta ,\;\;q_0\to -q_0)\;.
\label{eq:jdd}
\end{eqnarray}
The zeros of $D_\delta (q_0)$ correspond to poles of the mixed charged $%
\delta $-$\rho $ propagator. The pole contributions to the sum rules (\ref
{vec2}) are explicitly obtained from the expression
\begin{eqnarray}
\!\!\!\!\!\!\!\!\!{\rm sgn}(E_{j^{\pm }})\left| \langle j^{\pm }\mid J_0^{\pm }\mid C\rangle
\right| ^2 &=&-\left. \frac{m_u-m_d}{q_0-\rho }\frac{\left[ \delta J_{\delta
\delta }(q_0)-G_\delta ^{-1}(\delta -(m_u-m_d)/2)\right] }{dD_\delta
(q_0)/dq_0}\right| _{q_0=E_j^{\pm }}\;. \nonumber
 \\ &&
\label{eq:expolevec}
\end{eqnarray}


\end{document}